\documentclass[showpacs,twocolumn,superscriptaddress,floatfix,prd,authoryear,nofootinbib,aps,10pt,tightlines]{revtex4-2}
\usepackage{doi}
\usepackage{hyperref}
\hypersetup{
  colorlinks=true,        
  linkcolor=blue,         
  citecolor=cyan,         
}

\usepackage{graphicx}
\usepackage{dcolumn}
\usepackage{bm}
\usepackage{color}
\usepackage{enumitem}
\usepackage{amsmath}
\usepackage{amssymb}
\usepackage{orcidlink}
\usepackage{subcaption}
\usepackage{graphicx} 

\usepackage{float}
\usepackage{booktabs}
\usepackage{dcolumn,enumerate}
\usepackage{ragged2e}

\usepackage{xcolor}
\usepackage{caption}
\usepackage{appendix}
\usepackage{subcaption}

\newcommand {\be}{\begin{equation}}
\newcommand {\ee}{\end{equation}}
\newcommand {\bea}{\begin{array}}
	
	\newcommand {\eea}{\end{array}}

\begin{document}

\title{Epicyclic motion and accretion disk around a charged black hole in Einstein-ModMax theory with a quintessence field}
\author{Hamza Rehman}
\email{hamzarehman244@zjut.edu.cn}
\affiliation{Institute for Theoretical Physics and Cosmology, Zhejiang University of Technology, Hangzhou 310023, China}
\affiliation{United Center for Gravitational Wave Physics (UCGWP), Zhejiang University of Technology, Hangzhou 310023, China}
\affiliation{Center for Theoretical Physics, Khazar University, 41 Mehseti Str., Baku, AZ1096, Azerbaijan}

\author{Sanjar Shaymatov}
\email{sanjar@astrin.uz}
\affiliation{Institute of Fundamental and Applied Research, National Research University TIIAME, Kori Niyoziy 39, Tashkent 100000, Uzbekistan}
\affiliation{University of Tashkent for Applied Sciences, Str. Gavhar 1, Tashkent 100149, Uzbekistan}
\affiliation{Tashkent State Technical University, 100095 Tashkent, Uzbekistan}

\author{Tao Zhu}
\email{zhut05@zjut.edu.cn; Corresponding author}
\affiliation{Institute for Theoretical Physics and Cosmology, Zhejiang University of Technology, Hangzhou 310023, China}
\affiliation{United Center for Gravitational Wave Physics (UCGWP), Zhejiang University of Technology, Hangzhou 310023, China}

\date{\today}
\begin{abstract}

We investigate the epicyclic motion of charged test particles and the associated quasi-periodic oscillations (QPOs) around a weakly magnetized black hole surrounded by quintessence within the framework of Einstein–ModMax theory. We analyze the dynamics of charged particles on circular orbits and derive the corresponding radial and vertical epicyclic frequencies. The influence of the nonlinear electrodynamics parameter, magnetic coupling, dyonic charge, and quintessence state parameter on the innermost stable circular orbit and epicyclic frequencies is examined in detail. Using the forced resonance model, we compare the theoretical predictions of high-frequency QPOs with observational data from several X-ray binary systems. A Markov Chain Monte Carlo analysis is employed to constrain the black hole parameters and assess the role of weak magnetization and nonlinear electrodynamics effects. This analysis indicates that QPO observations tightly constrain the black hole mass and orbital radius while placing stringent upper bounds on the ModMax coupling, magnetic interaction, and dyonic charge. In addition, we study the radiative properties of the accretion disk and analyze the effects of the model parameters on the disk flux and temperature profiles. These findings suggest that the observed QPOs are consistent with general relativity in the strong-field regime, allowing only small deviations associated with Einstein–Maxwell theory in the presence of a quintessence field.
\end{abstract}
\pacs{
} \maketitle


\section{Introduction}
\label{introduction}

The general relativity (GR) theory of gravity is widely accepted, even though several issues remain unresolved. One of the important predictions of GR is the presence of black holes (BHs), which form when massive stars experience gravitational collapse at the end of their life cycle, that have remarkable gravitational properties  \cite{Christodoulou1986, Joshi1994, Joshi:2000fk, Goswami:2005fu, Harada:2001nj, Stuchlik:2012zza, Joshi2015}. Furthermore, significant progress has been achieved in 2019 due to the observations of the Event Horizon Telescope (EHT) after capturing the first horizon-scale image of the supermassive BH in the center of M87$^{*}$ galaxy \cite{EventHorizonTelescope:2019dse, EventHorizonTelescope:2019uob}. This achievement enables the researcher to compare the theoretical predictions of the BH shadow directly with observational evidence \cite{Yan:2025mlg, Gan:2021xdl, Liu:2020ola, Liu:2020vkh, Jusufi:2020cpn, Liu:2021yev, Zhu:2019ura, Jiang:2023img, Jiang:2024vgn, Shi:2024bpm, Uktamov:2024ckf}. Also, in 2022, the EHT captured the first image of the supermassive BH Sgr A$^{*}$ located at the core of the Milky Way galaxy, considerably increasing our understanding of these compact objects \cite{Lu:2018uiv, EventHorizonTelescope:2022wok, Grigorian:2024rsn}.

It is also well known that GR is unable to solve some problems, such as the lack of a consistent framework for spacetime quantization and the singularity problem, where GR is no longer applicable. In this respect, scientists have presented several alternative and modified theories of gravity that provide viable frameworks to address the problems of GR in strong gravitational fields. Thus, it becomes very important to understand the nature and remarkable aspects of the current gravitational fields, as well as how they affect test-particle geodesics in the surroundings of BHs. In this sense, current observations are crucial for understanding the geometric properties and fields surrounding the BHs. Therefore, these fields have become essential to examining the geodesic motion of particles in various physical contexts \cite{Shaymatov:2014fza, Shaymatov:2014dla, Dadhich:2018gmh, Shaymatov21pdu, Fatima:2025sdp}, as well as for influencing observable properties like the innermost stable circular orbit (ISCO) and quasi-periodic oscillations (QPOs) \cite{Wald:1974np, Benavides-Gallego:2018htf, Bini:2012zze, Toshmatov:2019bda, Dadhich22a}.
The QPOs are currently an important tool for studying the physics of strong-field gravity, as they can be detected through X-ray emission from BHs and neutron stars. The luminosity of oscillations is almost periodic, reflecting both relativistic gravitational phenomena and accretion dynamics \cite{Boshkayev:2020kle,Collodel_2021,Alloqulov24CPC,Alloqulov24EPJP,Chen2025,Nozari2025,Igata2025,XamidovAccretion2025}. Understanding how surrounding fields affect the geodesic motion of massive particles around a BH is essential. In particular, recent observations indicate that the accelerated expansion of the Universe can be explained by vacuum energy, modeled by the cosmological constant $\Lambda$, which introduces a repulsive effect in Einstein’s field equations~\cite{Cruz05,Stuchlik11,Shaymatov18a,Rayimbaev-Shaymatov21a,Giri23EPJP}.
Subsequently, quintessence was proposed as a dynamical matter field to explain the repulsive behavior of dark energy, providing an alternative to the cosmological constant~\cite{Peebles03,Caldwell09,Nozari:2020tks,Saghafi:2022pme}. Motivated by this idea, Kiselev derived a BH solution surrounded by quintessence, described by the equation of state $p = \omega_{q}\rho$, where the parameter $\omega_q$ lies in the range $(-1;-1/3)$ ~\cite{Kiselev2003aa, Hellerman2001JHEP}. The value $\omega_{q}=-1$ corresponds to the vacuum energy associated with the cosmological constant $\Lambda$, while $\omega_{q}=-1/3$ represents a different matter configuration. Guided by this motivation, throughout this paper, we vary the equation-of-state parameter $\omega$ within the range $(-1;-1/3)$ to study its effects on BH spacetimes.


The relationship between a BH and an external magnetic field can be modeled using two commonly used techniques in Einstein-Maxwell theory. The first technique employs the spacetime's Killing vectors to generate the corresponding vector potentials, using the external field as a perturbation of the spacetime \cite{Wald:1974np}. The second approach applies the Ernst transformation on a seed solution, usually from the Kerr-Newman-Taub-NUT family, to take into consideration the effect of the magnetic field on spacetime curvature \cite{Ernst:1976mzr, Siahaan:2021uqo, Ghezelbash:2021lcf}. The latter approach is very helpful for examining how null geodesics around a BH are altered by external magnetic fields. In recent decades, there has been an increasing interest in a specific type of nonlinear electrodynamics called modified Maxwell (ModMax) theory \cite{bandos2021p}. The numerous aspects of ModMax electrodynamics and those associated with BH solutions have been thoroughly studied in \cite{Sorokin:2021tge, Kosyakov:2020wxv, bandos2021modmax, Kruglov:2021bhs, Flores-Alfonso:2020euz, BallonBordo:2020jtw, Kubiznak:2022vft, Barrientos:2022bzm, Siahaan:2023gpc}. 

The investigation of nonlinear electrodynamics is interesting due to its potential to tackle problems related to field singularities. In ModMax theory, charges are effectively shielded by the nonlinear parameter using an exponential factor. It is noteworthy that the exact solution for a stationary charged BH in Einstein-ModMax theory is quite similar to the familiar Reissner-Nordstr\"{o}m solution \cite{Flores-Alfonso:2020euz}, allowing for the construction of dyonic BH solutions with both electric and magnetic charges. The characteristics of dyonic Einstein-ModMax BHs, such as their lensing properties, shadows, and quasinormal modes, have been examined in \cite{Pantig:2022gih}, suggesting that these phenomena could be potentially observed in future astronomical observations. According to these developments, it is natural to question the potential outcomes of investigating magnetization within the framework of Einstein-ModMax theory, particularly at the perturbative level as outlined by Wald \cite{Wald:1974np}. This technique is simpler than producing an exact magnetized BH solution in Einstein-ModMax theory \cite{Barrientos:2024umq} or using Ernst's method in the Einstein-Maxwell setup \cite{Ernst:1976mzr}. Applying Wald's method to create a magnetized, dyonic BH in Einstein-ModMax theory allows researchers to investigate how external magnetic fields and non-linear parameters affect the motion of charged timelike objects. The solution of the weakly magnetized BH in the Einstein-ModMax theory and the movements of a test charged timelike object affected by external magnetic fields are discussed in the \cite{Siahaan:2024ioa}.

In this work, QPOs are studied along with the radiative properties of the accretion disk surrounding the weakly magnetized BH surrounded by quintessence within the framework of Einstein-ModMax theory \cite{Siahaan:2024ioa}. Numerous theoretical models have been proposed to interpret QPOs, such as the forced resonance (FR) models, relativistic precession (RP) model, epicyclic resonance (ER) model, warped disk (WD) model, and parametric resonance (PR) models \cite{Stella:1997tc, Cadez:2008iv, Kostic:2009hp, Germana:2009ce, Kluzniak:2002bb, Abramowicz:2003xy, Rebusco:2004ba, Nowak:1996hg, Torok:2010rk, Torok:2011qy, Kotrlova:2020pqy, Shaymatov23ApJ, Xamidov25EPJC...85.1193X}. The current research focuses on the forced resonance model, where the upper and lower QPO frequencies are given by \cite{Kluzniak:2002bb, Banerjee:2022ffu} $\omega_{\rm UP} = \omega_{\theta} + \omega_{r}, \omega_{\rm L} = \omega_{\theta}$. The QPOs are among the most important phenomena because they originate from matter accreting close to compact objects, typically within a few gravitational radii, and carry information about relativistic effects in strong gravitational fields. Initially, numerous studies focused on high-frequency QPOs in neutron-star systems, but these investigations were later extended to stellar-mass and supermassive BHs \cite{PhysRevLett.82.17}. Consequently, QPO analyses have been employed to test the no-hair theorem and to examine possible deviations from the Kerr geometry in various scenarios, including modified-gravity theories, nonlinear electrodynamics, wormholes, and BH candidates such as GRO~J1655-40 \cite{Allahyari:2021bsq, Banerjee:2022chn, Bambi:2012pa, Bambi:2013fea, Deligianni:2021ecz, Deligianni:2021hwt, Maselli:2014fca, Wang:2021gtd, Jiang:2021ajk, Ashraf:2025lxs, Yang:2025aro, Guo:2025zca, Yang:2024mro, Liu:2023ggz, DeFalco:2023kqy, Bambi:2022dtw, Liu:2023vfh, Rehman:2025hfd, Azreg-Ainou:2020bfl, Liu:2023vfh, Xamidov25PDU, Rehman:2025knv, Liu:2023ggz, Xamidov:2025}. Also, numerous studies have been done on the motion of test particles and the resulting epicyclic frequencies in different BH spacetimes \cite{Borah:2025crf, Shaymatov:2023rgb, Stuchlik:2015sno, Banerjee:2022ffu, Rehman:2025knv}.
Furthermore, the analysis of epicyclic motion and its applications to QPOs associated with X-ray data from accretion disks around compact objects is essential, as it provides detailed information about the geometry of the compact object \cite{Bambi:2012ku, Bambi:2015ldr, Tripathi:2019bya}. For instance, high-frequency QPO models have been considered to address key characteristics of the epicyclic dynamics of neutral and charged particles in BH accretion disks \cite{Abramowicz:2004je, stuchlik2013multi, stella1999correlations, rezzolla2003new}.

The paper is organized as follows. In Sec.~\ref{Sec:metic}, we describe the spacetime of a weakly magnetized BH in Einstein-ModMax theory and study the motion of charged test particles. In Sec.\ref{Sec:qp0}, we analyze the epicyclic motions and compute the corresponding fundamental frequencies. In Sec .~\ref{Sec:parameter}, using observational data, we compare the epicyclic frequencies with the observed high-frequency QPOs and constrain the black-hole parameters. In Sec.~\ref{Sec:Acc}, we study the properties of the accretion disk around the considered BH. Finally, we conclude our work in Sec.~\ref{Sec:Conclusion}.

\section{Weakly magnetized ModMax BH and charged test particles dynamics \label{Sec:metic} }

The ModMax theory \cite{Kosyakov:2020wxv, Bandos:2020hgy} is described by Lagrangian density
\be 
{\cal L}_{\rm MM}  =  - \frac{1}{2}\left( {s\cosh v - \sqrt {s^2  + p^2 } \sinh v} \right),
\ee 
where $s$ and $p$ are the invariants of the electromagnetic fields, namely
\be \label{eq.sp}
s = \frac{1}{2}F_{\mu \nu } F^{\mu \nu} ~~,~~ p = \frac{1}{2}F_{\mu \nu } \tilde F^{\mu \nu } \,.
\ee 
In the equations above, the field strength tensor is defined as $F_{\mu \nu }  = \partial _\mu  A_\nu   - \partial _\nu  A_\mu  $ and its dual as $\tilde F_{\mu \nu }  = \frac{1}{2}\varepsilon _{\mu \nu \alpha \beta } F^{\alpha \beta } $ where ${\varepsilon _{0123} } = \sqrt { - g} $ with $g$ being the determinant of the metric tensor $g_{\mu\nu}$. In differential form notation, we have ${\bf {\tilde F}} = \star {\bf  F}$, where $\star$ denotes the Hodge dual star operator. The parameter $v$ denotes the non-linear parameter of the theory, with the standard Maxwell theory recovered when $v=0$. It has been shown that the condition $v \ge 0$ must be imposed to ensure causality \cite{Sorokin:2021tge}. Furthermore, since the ordinary Maxwell theory describes our nature extremely well, we can expect the parameter $v$ to be extremely small. 

Following \cite{BallonBordo:2020jtw, Barrientos:2022bzm}, we introduce the two-form for the ``material" field strength as follows
\be \label{eq.TensorMaterial}
{\bf E} = 2\left( {f_s {\bf F} + f_p {\bf {\tilde F}}} \right),
\ee 
where ${\bf F} $ and ${\bf {\tilde F}}$ are the two-forms for the Maxwell field-strength tensor and its dual, respectively. The functions $f_s$ and $f_p$ are defined as
\begin{eqnarray} 
f_s  = \frac{{\partial {\cal L}_{MM} }}{{\partial s}} \mbox{~~and~~} f_p  = \frac{{\partial {\cal L}_{MM} }}{{\partial p}} \, .
\end{eqnarray} 
In terms of the functions above, the electromagnetic stress-energy tensor in ModMax theory can be written as
\be \label{eq.Tmn}
T_{\mu \nu }  = \frac{1}{{4\pi }}\left( {s~ g_{\mu \nu }  - 2 F_{\mu \kappa } F_{\nu \lambda } g^{\kappa \lambda } } \right) f_s \, .
\ee 
Furthermore, the electric charge inside a closed two-dimensional spacelike surface $\Sigma $ can be obtained from the integral
\be \label{eq.Qe}
Q_e  = \frac{1}{{4\pi }}\int_\Sigma  {  \star {\bf  E}} \,.
\ee 

Now let us consider an action for the Einstein-ModMax theory \cite{Flores-Alfonso:2020euz, BallonBordo:2020jtw, Barrientos:2022bzm}
\be 
S = \frac{1}{{16\pi }}\int {d^4 x\sqrt { - g} } \left( {R - 4{\cal L}_{MM} } \right).
\ee 
From the action above, the corresponding equations of motion in Einstein-ModMax theory are
\be \label{eq.EinsteinEq}
R_{\mu \nu }  - \frac{1}{2}g_{\mu \nu }  = 8\pi T_{\mu \nu } \,
\ee 
and
\be \label{eq.source-free}
\nabla ^\mu  E_{\mu \nu }  = 0\, .
\ee 
where, \(E^{\mu\nu}\) indicates the constitutive tensor of ModMax electrodynamics, as given in Eq. \ref{eq.TensorMaterial}.
This tensor generalizes the Maxwell field tensor and reduces to it \(F^{\mu\nu}\) in the linear limit.
The above equation represents the generalized source-free condition,
which replaces \(\nabla^\mu F_{\mu\nu}=0\) in standard Einstein--Maxwell theory.

Despite the complexity of the corresponding equations of motion, it turns out that one of the simplest static BH solutions describing a collapsed charged mass closely resembles the well-known Reissner-Nordstr\"{o}m solution. The spacetime metric describing a weakly magnetized BH surrounded by the quintessential field in Einstein-ModMax theory reads \cite{Kiselev2003aa,Flores-Alfonso:2020euz} 
\be \label{eq.Metric}
ds^2  =  - f(r)dt^2  + \frac{{dr^2 }}{f(r)} + r^2 \left( d\theta^2 + \sin^2\theta d\phi ^2  \right)\, ,
\ee 
where 
$$f(r)=\left( {1 - \frac{{2M}}{r} + \frac{{e^{ - \nu} Q^2 }}{{r^2 }}-\frac{c}{ r^{1+3 \omega }}} \right)\, ,$$
whereas the corresponding vector potential is given by
\be \label{eq.Anonmag}
A_\mu  dx^\mu   = -\frac{{e^{ - v} Q}}{r}dt\, .
\ee 
In equations above, $M$ is the BH mass and $Q=\sqrt{Q^2_{e}+Q^2_{m}}$ is the BH charge with $Q_e$ and $Q_m$ being the electric and magnetic charges respectively, whereas $\nu$ denotes the non-linear parameter of the ModMax theory. The parameters $\omega$ and $c$ respectively denote the equation of state of the quintessential field and the quintessential field parameter that refers to the intensity of the quintessence energy field.

For the solution described by Eqs.~(\ref{eq.Metric}) and (\ref{eq.Anonmag}), the non-vanishing components of the field strength tensor and its dual are given by ${\bf F} = -\frac{{e^{-\nu} Q}}{{r^2 }}dt \wedge dr$
and ${\bf \tilde F} = e^{ -\nu} Q\sin \theta dt \wedge dr$, respectively. Accordingly, by using Eq.~(\ref{eq.sp}) one can find 
\begin{eqnarray} 
s =  - \frac{{e^{ - 2\nu} Q^2 }}{{2r^4 }} \mbox{~~and~~} p=0 .
\end{eqnarray} 

We investigate the dynamics of a charged test particle with electric charge $q$ and mass $m$. The Hamiltonian governing the system is given by \cite{C.W.Misner} 
\begin{eqnarray}
\mathcal{H} \equiv \frac{1}{2} g^{\alpha\beta} \left( \frac{\partial \mathcal{S}}{\partial x^{\alpha}} - q A_{\alpha}\right)
\left( \frac{\partial \mathcal{S}}{\partial x^{\beta}} - q A_{\beta} \right),
\end{eqnarray}
where $\mathcal{S}$ denotes the action, $x^{\alpha}$ refers to the spacetime four-vector coordinates, and
\begin{eqnarray}
A_{\alpha} = \left( -\frac{e^{-\nu} Q_e}{r},\, 0,\, 0,\, Q_m (1 - \cos\theta)\right)\, , 
\end{eqnarray}
are the non-zero components of the electromagnetic four-potential.

Following this equation, we will examine the Hamiltonian of the system, i.e., $H = k/2$, where $k = -m^2$. The Hamilton–Jacobi action $S$ is expressed as
\begin{eqnarray}
S = -\frac{1}{2}k\lambda - Et + L\varphi + S_r(r) + S_\theta(\theta), \label{1a}
\end{eqnarray}
were $S_r$ and $S_\theta$ are the functions of $r$ and $\theta$, respectively. From Eq.~(\ref{1a}), we have 
\begin{eqnarray}
&&-\frac{1}{f(r)}\left(\frac{e^{-\nu } q Q_e}{r}-E\right)^2 + f(r)\left(\frac{\partial S_r}{\partial r} \right)^2 + \left(\frac{1}{r} \frac{\partial S_\theta}{\partial \theta} \right)^2
 \nonumber \\
&&+\frac{[L-q Q_m (1-\cos\theta)]^2}{r^2 \sin^2 \theta}-k=0 . \label{2}
\end{eqnarray}
From the above Hamilton–Jacobi equation, we see that there are four conserved quantities: the specific energy, the angular momentum $k$~\cite{C.W.Misner}, and a fourth associated with the latitudinal motion, which can be neglected for particles moving in the equatorial plane (i.e., $\theta = \pi/2$). For convenience, we split Eq.~(\ref{2}) into two parts: the dynamical part $H_{\text{dyn}}$ and the potential part $H_{\text{pot}}$, as follows.
\begin{eqnarray}\label{3}
H_{\text{dyn}} &=& \frac{1}{2} \left[ 
f(r) \left( \frac{\partial S_r}{\partial r} \right)^2 
+ \frac{1}{r^2} \left( \frac{\partial S_\theta}{\partial \theta} \right)^2 
\right]\, ,\\ 
H_{\text{pot}} &=& \bigg[ 
- \frac{\left( \frac{e^{-\nu} q Q_e}{r} - E \right)^2}{f(r)} - k \nonumber\\ 
&& + \frac{(L-q Q_m (1-\cos\theta))^2 }{r^2 \sin^2\theta}\bigg]\, . \label{4}
\end{eqnarray}
The angular $S_\theta$ and radial parts $S_r$ of the Hamiltonian given by
\begin{eqnarray}
S_\theta &=& \int \sqrt{ K - \left( \frac{L-q Q_m (1-\cos\theta)}{\sin \theta} \right)^2 } \, d\theta, \\
S_r &=& \int \sqrt{ \Big(\frac { e^{-\nu} q Q_e}{r}- E\Big)^2 - f(r)\Big( -k + \frac{K}{r^2} \Big)} \frac{dr}{f(r)}. \nonumber \\
\end{eqnarray}
For subsequent investigation, we define
\begin{eqnarray}
\mathcal{E} = \frac{E}{m}, \quad  \mathcal{K} = \frac{K}{(mM)^2}, \quad \frac{k}{m^2} = -1, \quad \mathcal{L} = \frac{L}{mM},\label{5}
\end{eqnarray}
and the radial coordinate is normalized as $r\rightarrow{r/M}$. Moreover, we compute the following equation of motion by using the Hamiltonian system 
\begin{eqnarray}
\dot t &=& \frac{1}{f(r)} \left( \mathcal{E} - \frac{g_{e}e^{-v}}{r} \right)\, , \\ \label{6a}
\dot \phi &=& \frac{\left( \mathcal{L} - \sigma_{m}(1 - \cos\theta) \right)}{r^2 \sin^2\theta}\, , \\ \label{7}
\dot r &=& \sqrt{
\left( \mathcal{E} - \frac{g_{e} e^{-v}}{r} \right)^2 
- f(r) \left( 1 + \frac{\mathcal{K}}{r^2} \right)
}\, ,\\ \label{8}
\dot \theta &=& \frac{1}{r^2} \sqrt{
\mathcal{K} - \frac{ \left( \mathcal{L} - \sigma_{m} (1 - \cos\theta) \right)^2 }{\sin^2\theta}} ,  \label{9}
\end{eqnarray}
where $g_{e}=\frac{q Q_e}{mM}$, $\sigma_{m}=\frac{q Q_m}{mM}$, and the dot represents the derivative w.r.t proper time $\tau$. For the radial motion of the charged particles, the effective potential is 
\begin{eqnarray}
V_{\mathrm{eff}}(r) = \frac{g_{e}\, e^{-\nu}}{r} \;+\; \sqrt{\, f(r) \left( 1 + \frac{\mathcal{L}^{2}}{r^{2}} \right) }\, . \label{10}
\end{eqnarray}
\begin{figure*}
\centering
\centering
a.\includegraphics[scale=0.25]{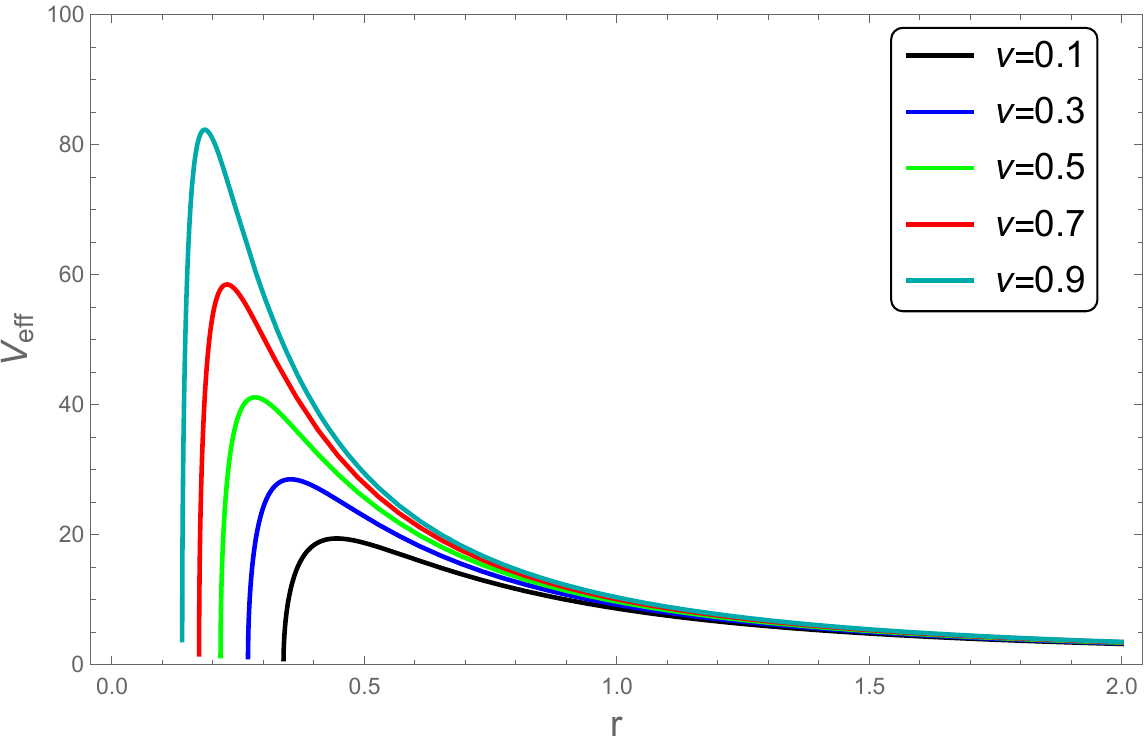} b.\includegraphics[scale= 0.25]{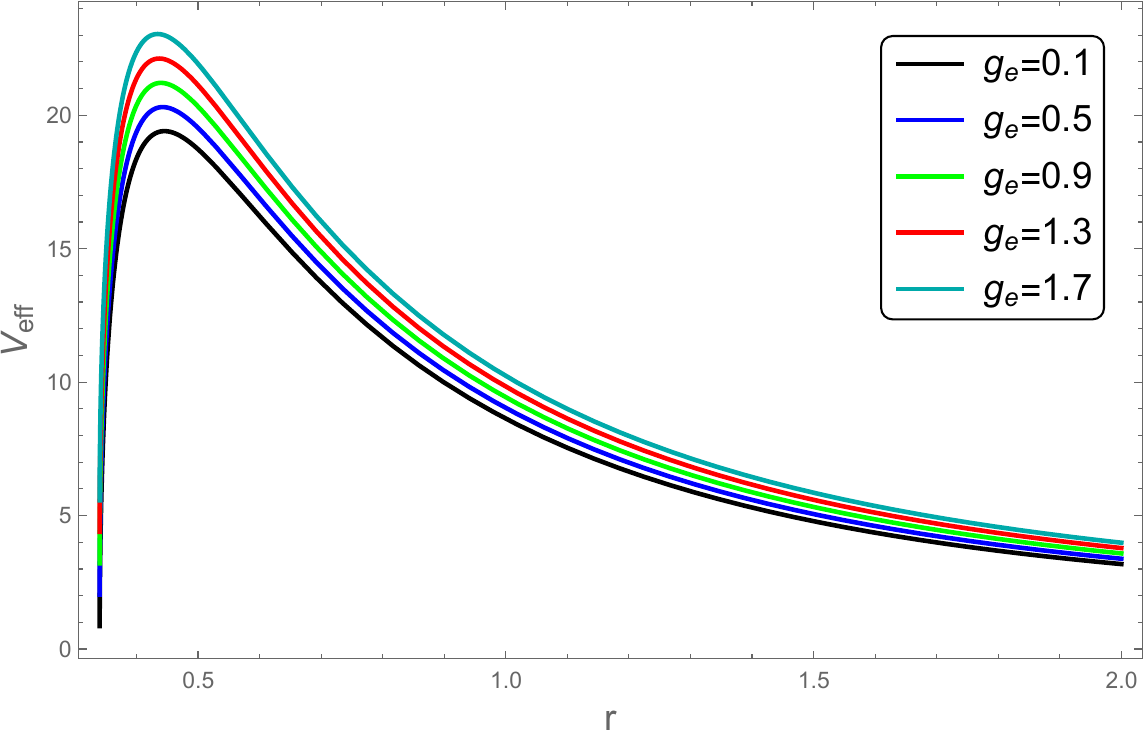} c.\includegraphics[scale= 0.25]{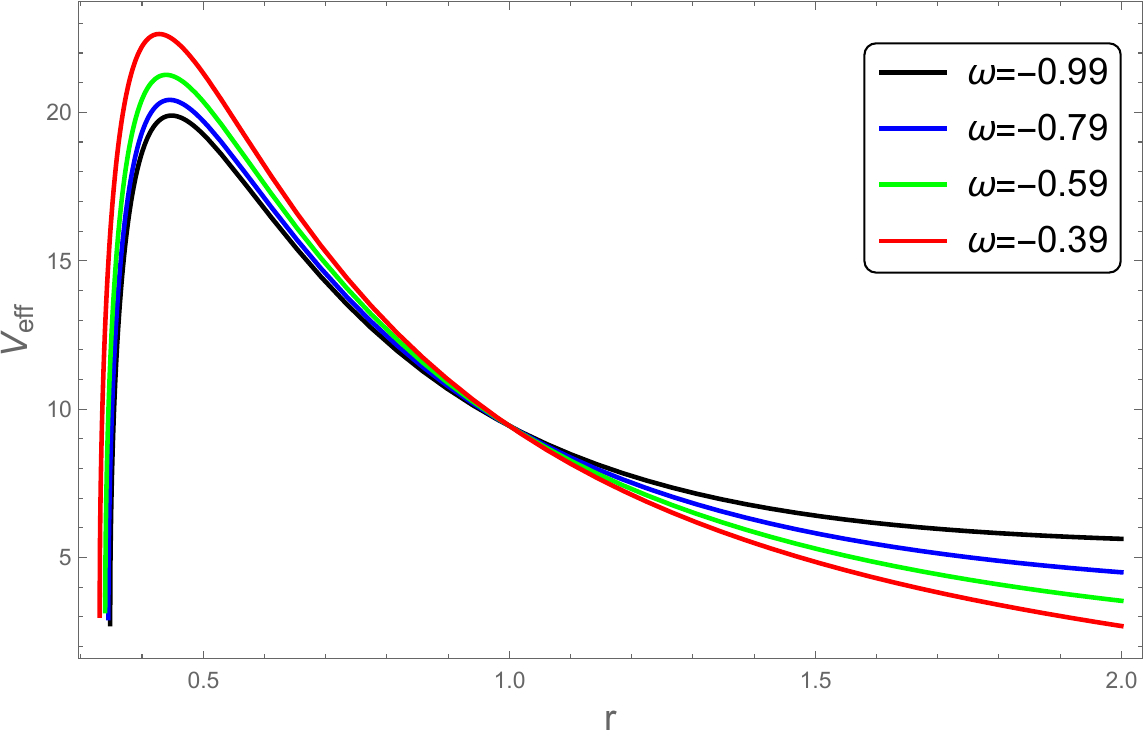}
\caption{Effective potential $V_{\mathrm{eff}}$ as a function of radius $r$ for test-particle radial motion with black-hole parameters $M=1$, $c=0.3$, and $\mathcal{L}=10$. 
(a) For $Q/M=0.8$, $\omega=-0.6$, $c=0.3$, $g_{e}=0.1$, varying $\nu$; 
(b) for $Q/M=0.8$, $\omega=-0.6$, $c=0.3$, $\nu=0.1$, different values of  $g_{e}$; 
(c) for $Q/M=0.8$, $g_{e}=0.1$, $c=0.3$, $\nu=0.1$, altered values of $\omega$.}
\label{fig:1}
\end{figure*}

The graphical analysis of the effective potential for the motion of charged particles along $r$ is depicted in Fig.~\ref{fig:1}. It is worth noting that the maximum value of the effective potential is observed in the vicinity of the BH horizon and declines as one moves away from it. From Fig.~\ref{fig:1} a, it is clear that the effective potential rises with increasing values of the ModMax parameter $\nu$; and similar behavior is obtained for the parameter $g_{e}$, as shown in Fig.~\ref{fig:1} b. However, when particles are far from the BH, the effective potential declines due to the increase in the quintessence state parameter $\omega$, while the reverse behavior is observed in the vicinity of the BH (see Fig.~\ref{fig:1}c). 

We now aim to determine the particle's motion in circular orbits in the vicinity of the considered BH. For such orbits to exist, the particles must satisfy the conditions
\begin{eqnarray}
    V_{\rm eff}=\mathcal{E}, \,\,\,\,\,\,\,\, \frac{V_{\rm eff}}{dr} =0. \label{11}
\end{eqnarray}
From the above expression, we get the following result
\begin{eqnarray}
 \mathcal{L}^2 &=&
\frac{2 g_e^2 F + 2 r^3 F F_{,r} - r^4 F_{,r}^2}
     {r^2 4 F^2 - 4 r F F_{,r} + F_{,r}^2}
+ \nonumber\\&& \frac{2 \sqrt{-2 g_e^2 r^3 F^2 F_{,r} + g_e^4 F^2 + 4 g_e^2 r^2 F^3}}
       {4 F^2 - 4 r F F_{,r} + r^2 F_{,r}^2}. \label{12}   
\end{eqnarray}
It is important to note that circular orbits exist at the extrema (minima and maxima) of the effective potential for particles with angular momentum $\mathcal{L}$. The stable circular orbit exists at local minima of the effective potential, while the unstable circular orbit corresponds to the local maxima of the effective potential. The transition between these regions is characterized by the ISCO, defined as the radius at which the maxima and minima of $V_{eff}$ coincide at a single inflection point. Therefore, computing the ISCO radius is essential for understanding the dynamics of particles in its vicinity, which is determined by Eq. (\ref{11}) with one additional condition
\begin{eqnarray}
\frac{d^2V_{\rm eff}}{dr^2}=0\, .
\end{eqnarray}
We compute $\mathcal{E}_{\mathrm{ISCO}}$, $\mathcal{L}_{\mathrm{ISCO}}$, and $r_{\mathrm{ISCO}}$ numerically. The corresponding numerical values are presented in Table~\ref{tab:ISCOvalues}. From Table~\ref{tab:ISCOvalues}, it is evident that the ISCO radius decreases with increasing electromagnetic charge of the BH. This reduction is accompanied by decreases in $\mathcal{L}_{\rm ISCO}$ and $\mathcal{E}_{\rm ISCO}$, indicating that the electromagnetic charge enhances the gravitational attraction and permits stable circular orbits closer to the central object.  Also, the influence of the coupling parameter $g_{e}$ is observed from Table~\ref{tab:ISCOvalues}.  It is noted that for the case $g_{e}>0$, the ISCO radius shifts outward because the repulsive effect of the $ g_{e}$ counteracts electromagnetic charge and pushes the ISCO radius outward. Although for the scenario $g_{e}<0$, compressing the ISCO radius closer to the BH because the negative values of $g_{e}$ contribute to enhancing the gravitational force. 
\begin{table*}
\caption{The numerical values of $r_{\mathrm{ISCO}}$, $\mathcal{E}_{\mathrm{ISCO}}$, $\mathcal{L}_{\mathrm{ISCO}}$, $\Omega_{\mathrm{ISCO}}$, and $v_{\mathrm{ISCO}}$ for $M=1$ and $c=0.000005$, $\omega=-0.5$, $\nu=0.13$ shown for different values of $Q/M$ with $g_{e}=\pm 0.2$ and $g_{e}=\pm 0.4$.}
\label{tab:ISCOvalues}
\begin{ruledtabular}
\begin{tabular}{c|ccccc|ccccc}
$Q/M$ & \multicolumn{5}{c|}{$g_{e} = 0.2$} & \multicolumn{5}{c}{$g_{e} = 0.4$} \\ \hline
 & $\mathcal{L}_{ISCO}$ & $\mathcal{E}_{ISCO}$ & $r_{ISCO}$ & $v_{ISCO}$ & $\Omega_{ISCO}$ 
 & $\mathcal{L}_{ISCO}$ & $\mathcal{E}_{ISCO}$ & $r_{ISCO}$ & $v_{ISCO}$ & $\Omega_{ISCO}$ \\ \hline
0.1 & 3.124348 & 0.954049 & 6.010695 & 0.499247 & 0.067806 
    & 2.766638 & 0.965934 & 6.112577 & 0.493051 & 0.066118 \\
0.2 & 3.111051 & 0.953734 & 5.966065 & 0.501624 & 0.068412 
    & 2.752314 & 0.965678 & 6.064126 & 0.495226 & 0.066782 \\
0.3 & 3.088567 & 0.953193 & 5.892112 & 0.502842 & 0.069431 
    & 2.734097 & 0.965237 & 5.982220 & 0.497220 & 0.067875 \\
0.4 & 3.056377 & 0.952398 & 5.785569 & 0.506176 & 0.070951 
    & 2.697930 & 0.964585 & 5.864974 & 0.501104 & 0.069527 \\
\end{tabular}

\begin{tabular}{c|ccccc|ccccc}
$Q/M$ & \multicolumn{5}{c|}{$g_{e} = -0.2$} & \multicolumn{5}{c}{$g_{e} = -0.4$} \\ \hline
 & $\mathcal{L}_{ISCO}$ & $\mathcal{E}_{ISCO}$ & $r_{ISCO}$ & $v_{ISCO}$ & $\Omega_{ISCO}$ 
 & $\mathcal{L}_{ISCO}$ & $\mathcal{E}_{ISCO}$ & $r_{ISCO}$ & $v_{ISCO}$ & $\Omega_{ISCO}$ \\ \hline
0.1 & 3.779312 & 0.931799 & 6.004425 & 0.499636 & 0.067912 
    & 4.087360 & 0.921448 & 6.046050 & 0.497070 & 0.067212 \\
0.2 & 3.766272 & 0.931399 & 5.966545 & 0.500638 & 0.068418 
    & 4.073163 & 0.921017 & 6.009680 & 0.498306 & 0.067661 \\
0.3 & 3.744253 & 0.930714 & 5.894953 & 0.502331 & 0.069289 
    & 4.050630 & 0.920270 & 5.934400 & 0.499649 & 0.068468 \\
0.4 & 3.712793 & 0.929712 & 5.806684 & 0.504812 & 0.070568 
    & 4.018293 & 0.919289 & 5.857585 & 0.501570 & 0.069418 \\
\end{tabular}
\end{ruledtabular}
\end{table*}
\begin{table*}
\caption{The numerical values of $r_{\mathrm{ISCO}}$, $\mathcal{E}_{\mathrm{ISCO}}$, $\mathcal{L}_{\mathrm{ISCO}}$, $\Omega_{\mathrm{ISCO}}$, and $v_{\mathrm{ISCO}}$ for fixed $Q/M=0.3$, $g_{e}=0.2$, $M=1.0$, and $c=5\times10^{-6}$. The left panel corresponds to fixed $\nu=1.5$ with varying $\omega$, while the right panel corresponds to fixed $\omega=-0.50$ with varying $\nu$.}
\label{tab:ISCO_nu_w_split}
\begin{ruledtabular}
\begin{tabular}{c|ccccc||c|ccccc}
\multicolumn{6}{c||}{$\nu=1.5$} & \multicolumn{6}{c}{$\omega=-0.50$} \\ \hline
 $\omega$ & $r_{ISCO}$ & $\mathcal{E}_{ISCO}$ & $\mathcal{L}_{ISCO}$ & $\Omega_{ISCO}$ & $v_{ISCO}$ &
 $\nu$ & $r_{ISCO}$ & $\mathcal{E}_{ISCO}$ & $\mathcal{L}_{ISCO}$ & $\Omega_{ISCO}$ & $v_{ISCO}$ \\ \hline
 -0.40 &  5.992210 & 0.953946 & 3.119089 & 0.068060 & 0.499440 &
 0.00 & 5.877356 & 0.953107 & 3.084371 & 0.069642 & 0.502951 \\
 -0.6 & 5.992511 & 0.953932 & 3.119025 & 0.068053 & 0.499415 &
 0.05 & 5.884662 & 0.953161 & 3.086579 & 0.069539 & 0.502724 \\
 -0.8 & 5.994414 & 0.953882 & 3.118665 & 0.068013 & 0.499260&
 0.10 & 5.891601 & 0.953212 & 3.088675 & 0.069442 & 0.502509 \\
\end{tabular}
\end{ruledtabular}
\end{table*}

Moreover, we have noted how the quintessence and ModMax parameters influence the behavior of the particles moving in the circular orbits in the vicinity of the BH, which is given in the Table~\ref{tab:ISCO_nu_w_split}. It is clear that the quintessence state parameter $\omega$ weakly influences the ISCO radius. As the state parameter $\omega$ increases from $-0.8$ to $-0.4$ and keeping $\nu=1.5$ the ISCO radius shifts to the BH horizon along with the marginal increments in the specific energy and specific angular momentum of the particles in ISCO orbits. Conversely, varying $\nu$ and considering $\omega=0.05$, the ISCO shifts significantly away from the horizon, accompanied by higher $\mathcal{E}_{ISCO}$ and $\mathcal{L}_{\rm ISCO}$.

Also, we have examined how the quintessence and ModMax parameters affect the motion of particles in circular orbits around the BH, as summarized in Table~\ref{tab:ISCO_nu_w_split}. The results show that the quintessence state parameter $\omega$ exerts only a weak influence on the ISCO radius. For fixed $\nu=1.5$, increasing $\omega$ from $-0.8$ to $-0.4$ produces a slight inward shift of the ISCO toward the BH horizon, accompanied by marginal increases in the $\mathcal{E}_{\rm ISCO}$ and $\mathcal{L}_{\rm ISCO}$. In contrast, varying the nonlinear ModMax parameter $\nu$ at fixed $\omega=-0.50$ leads to a more pronounced outward shift of the ISCO radius from the central object, together with higher values of $\mathcal{E}_{\rm ISCO}$ and $\mathcal{L}_{\rm ISCO}$.

Next, we extend our discussion to determine the radial and angular velocities of the particles moving on the ISCO in the vicinity of the weakly magnetized Einstein-ModMax BH. For this purpose, we use Eqs.~(\ref{6a}-\ref{9}), and introduce the coordinate velocity computed by the local observer \cite{tredcr1975cw,shapiro2024black,Shaymatov:2023rgb}. The components of coordinates velocity given by
\begin{eqnarray}
v_{\hat r}
&=& \sqrt{\,1 - \frac{f(r)}{\left(\mathcal{E}-\dfrac{g_{e}\,e^{-\nu}}{r}\right)^{2}}
\left(1+\frac{\mathcal{K}}{r^{2}}\right)}\, ,\label{13} \\
v_{\hat\theta}
&=& \frac{\sqrt{f(r)}}{\,r\!\left(\mathcal{E}-\dfrac{g_{e}\,e^{-\nu}}{r}\right)}\,
\sqrt{\,K-\frac{\big(L-\sigma_m(1-\cos\theta)\big)^{2}}{\sin^{2}\theta}}\,, \label{14} \nonumber \\
\;\; \\
v_{\hat\phi}
&=& \frac{\sqrt{f(r)}}{\,r\!\left(\mathcal{E}-\dfrac{g_{e}\,e^{-\nu}}{r}\right)}\,
\frac{L-\sigma_m(1-\cos\theta)}{\sin\theta}\, .\label{15}
\end{eqnarray}
The motion of the particles is restricted in the equatorial plane (i.e., $\theta=\pi/2$), therefore, $\mathcal{K}$ is neglected, and by using Eq.~(\ref{13}-\ref{15}), we attain
\begin{eqnarray}
\mathcal{E}
= \frac{\sqrt{f(r)}}{\sqrt{1 - v^{2}}}
+ \frac{g_{e}\,e^{-\nu}}{r}, \;\;\; 
v^{2} = v_{\hat r}^{2} + v_{\hat \theta}^{2} + v_{\hat \phi}^{2}.\label{16}
\end{eqnarray}
It is interesting to note that in the vicinity of the BH horizon where $f(r) \rightarrow{0}$, the radial component of velocity approaches the speed of light (i.e. $v_{\hat r}\rightarrow{1}$) and the remaining components $v_{\hat \theta}=0$ and $v_{\hat \phi}=0$. Determining the particle's linear velocity at the ISCO is essential because it precisely describes the dynamics of accreting matter in the intense gravitational field. Therefor, one can define the orbital velocity $v_{\phi}$ in Ref. \cite{Pugliese:2011py, Shaymatov:2021nff} is given by
\begin{eqnarray}
v = v_{\phi} = \Omega \sqrt{-\frac{g_{\phi\phi}}{g_{tt}}} ,\label{16a}
\end{eqnarray}
where $\Omega$ represents the orbital angular velocity of the particle, corresponding to the Keplerian angular frequency in the neutral case. At the ISCO, the radial and polar components of the velocity vanish, i.e., $v_r = v_\theta = 0$, so that only the azimuthal motion of the particles contributes to the dynamics. For charged particles, the orbital frequency $\Omega$ must be obtained from the non-geodesic equation of motion, which incorporates both gravitational and electromagnetic interactions. Generally, one can determine the angular velocity for the charge particles that follow the non-geodesic equation%
\begin{eqnarray}
g_{tt,r} + \Omega^{2} g_{\phi\phi,r} 
= -\frac{2q}{m} \, \frac{\Omega A_{\phi,r}}{\sqrt{-g_{tt} - \Omega^{2} g_{\phi\phi}}} \,. \label{17}
\end{eqnarray}
From the above expression, we get \cite{Shaymatov:2022enf}
\begin{eqnarray}
\Omega^{2} &=&
\Bigg\{ \Omega_{0}^{2}
 - 2 g_{tt}\Big(\frac{q A_{\phi,r}}{m g_{\phi\phi,r}}\Big)^{2}
 \pm \frac{2 q A_{\phi,r}}{m g_{\phi\phi,r}}  \nonumber \\
&& \times
\Bigg[ - g_{tt}\Omega_{0}^{2}
 - g_{\phi\phi}\Omega_{0}^{4}
 + \Big( \frac{q A_{\phi,r}\, g_{tt}}{m g_{\phi\phi,r}} \Big)^{2} \Bigg]^{1/2}\Bigg\} \nonumber \\
&& \times
\left[ 1 + 4 g_{\phi\phi}
  \Big(\frac{q A_{\phi,r}}{m g_{\phi\phi,r}}\Big)^{2} \right]^{-1}.\label{18}
\end{eqnarray}
The above expression reduces to $\Omega^{2} = \Omega_{0}^{2} = -\frac{g_{tt,r}}{g_{\phi\phi,r}}$ if $q=0$. From Eq.~(\ref{18}) we have
\begin{eqnarray}
\Omega_{k}=\sqrt{-\frac{\left[c (-3 \omega -1) r^{-3 \omega -2}-\frac{2 M}{r^2}+\frac{2 e^{-\nu } Q^2}{r^3}\right]}{2r\sin ^2(\theta )}}\, . \label{19}
\end{eqnarray}
 From Eq. (\ref{16a}), we obtained the orbital velocity at the equatorial plane
\begin{eqnarray}
v= \sqrt{\frac{2 r^{3 \omega } \left(Q^2-e^{\nu } M r\right)-c e^{\nu } r (3 \omega +1)}{2 \left[c e^{\nu } r-r^{3 \omega } \left(e^{\nu } r (r-2 M)+Q^2\right)\right]}}. \label{20}
\end{eqnarray}
\begin{figure*}
\centering
a. \includegraphics[scale= 0.28]{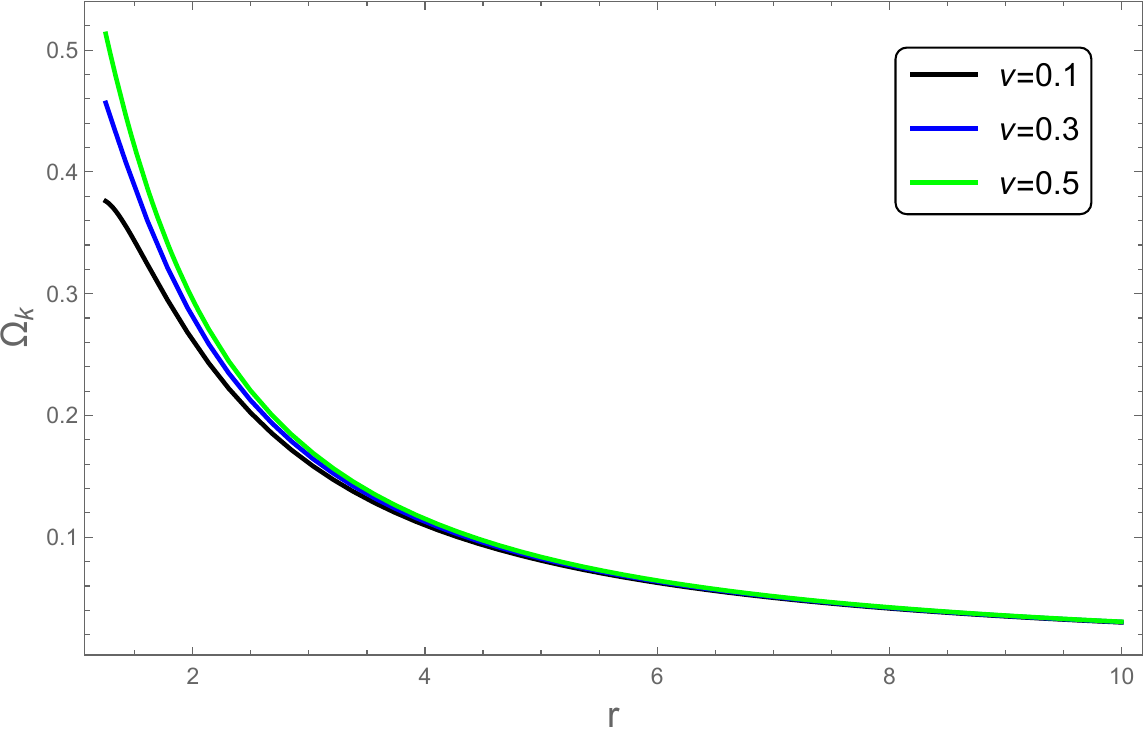}
b. \includegraphics[scale=0.28]{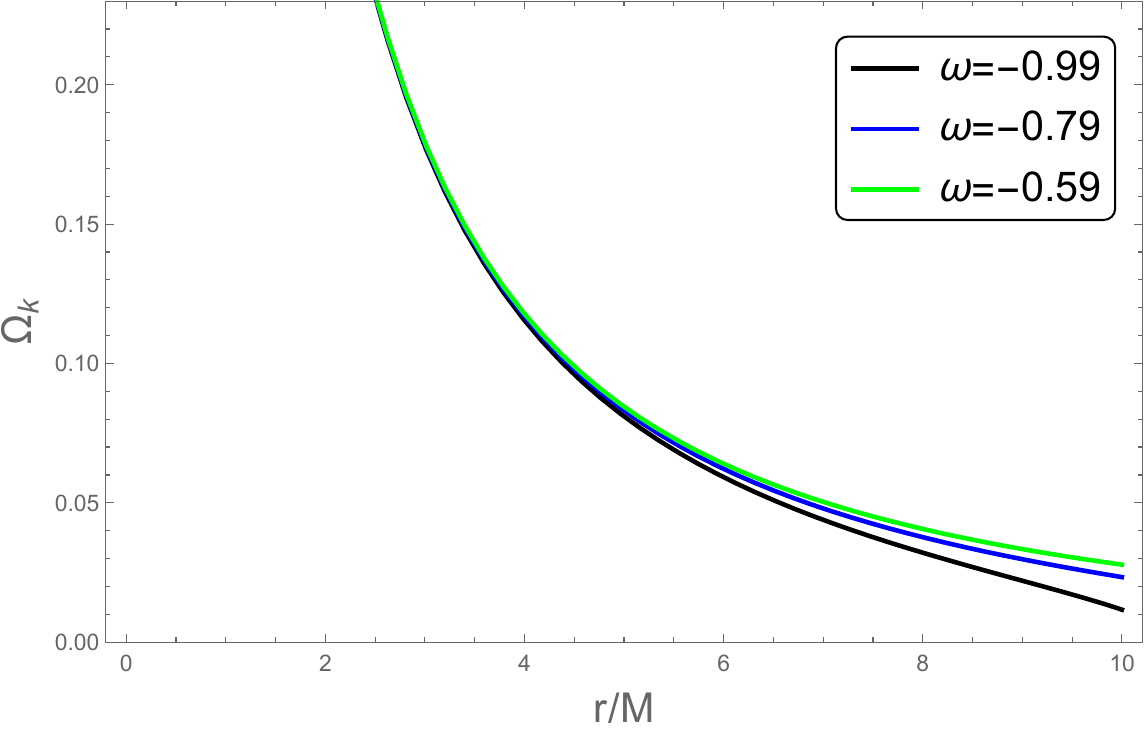}
c. \includegraphics[scale= 0.28]{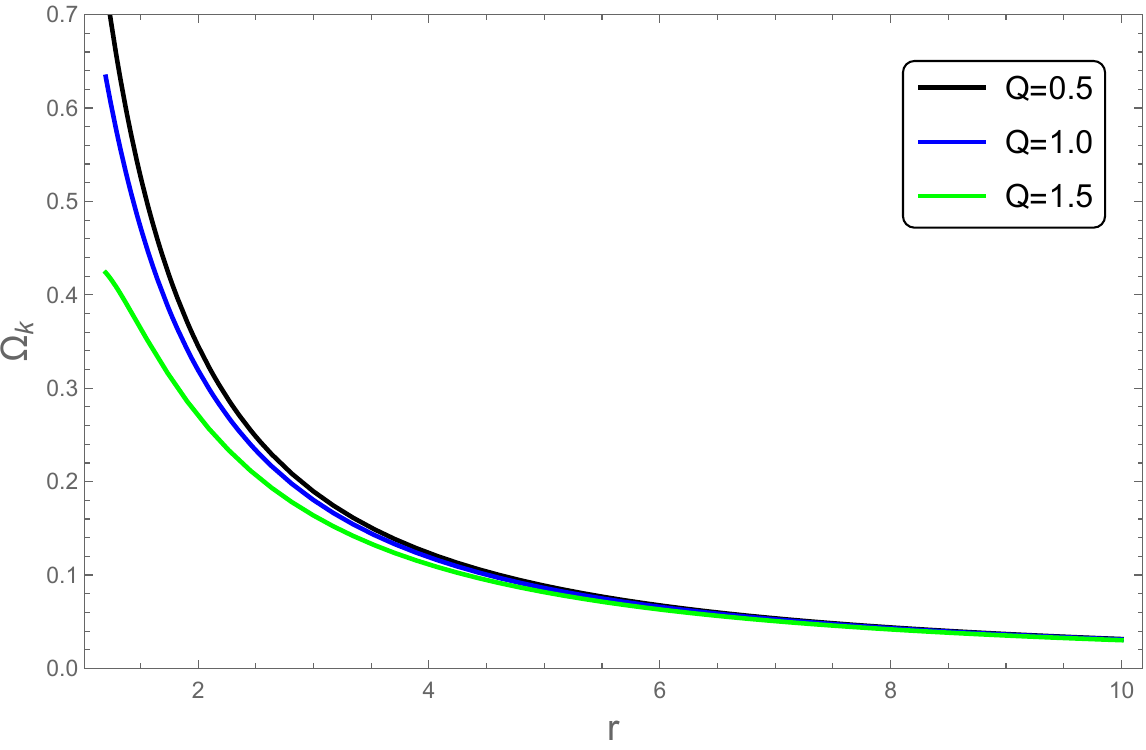}
\caption{The Keplerian frequency $\Omega_{k}$ is plotted along $r/M$ for $M=1$, $c=0.0009$: 
a. $Q/M=1$, $\omega =-0.39$ and various values of  $\nu$; 
b. $\nu=1$, $Q=1$, and different values of $\omega$;
c. $\omega=-0.39$, $\nu=1$ and several values of $Q/M$.}\label{fig:2}
\end{figure*}
\begin{figure*}
\centering
a. \includegraphics[scale= 0.28]{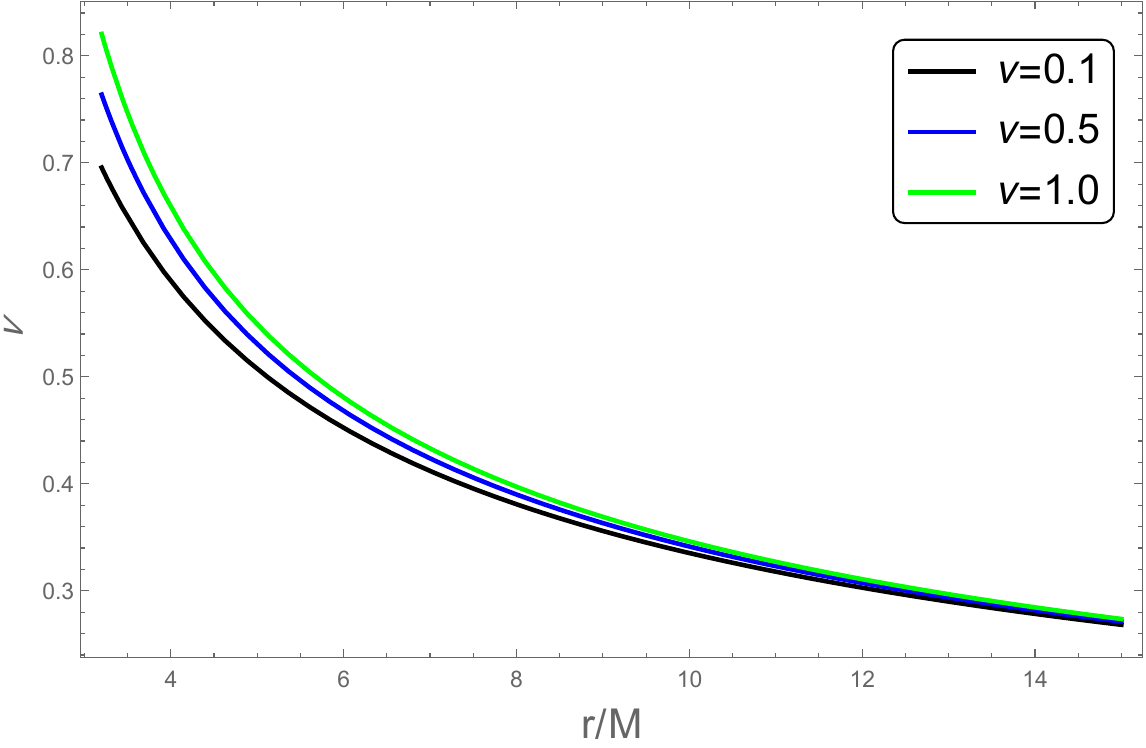}
b. \includegraphics[scale= 0.28]{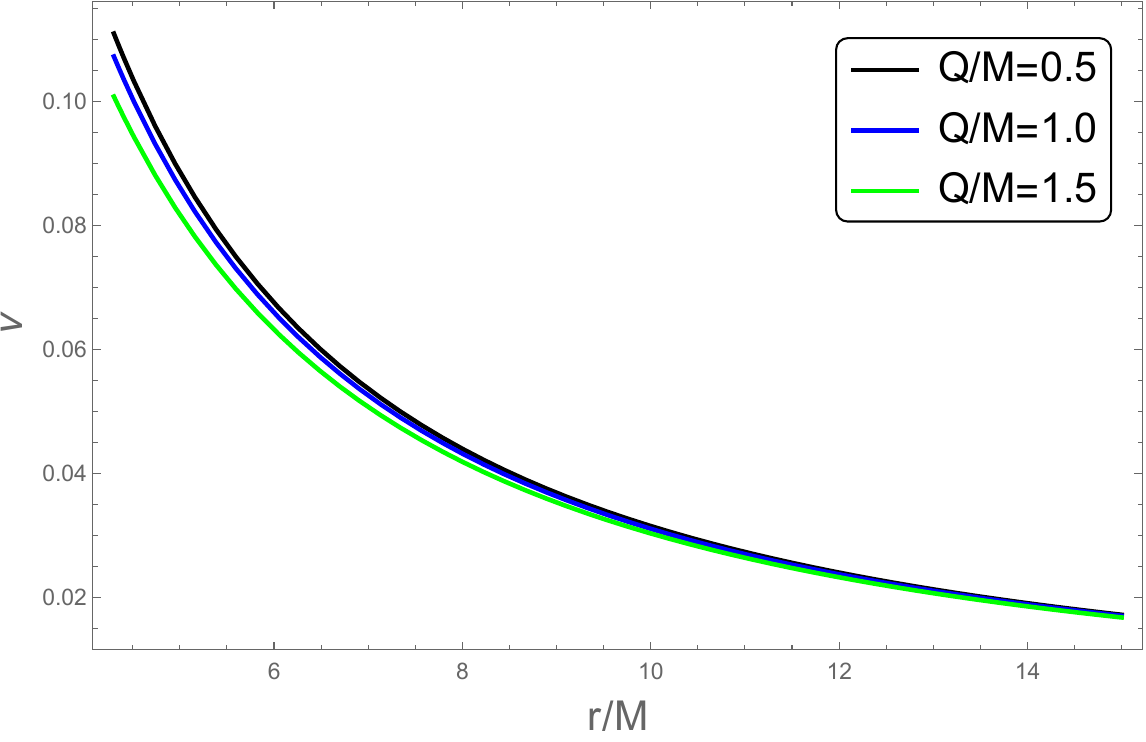}
c. \includegraphics[scale= 0.28]{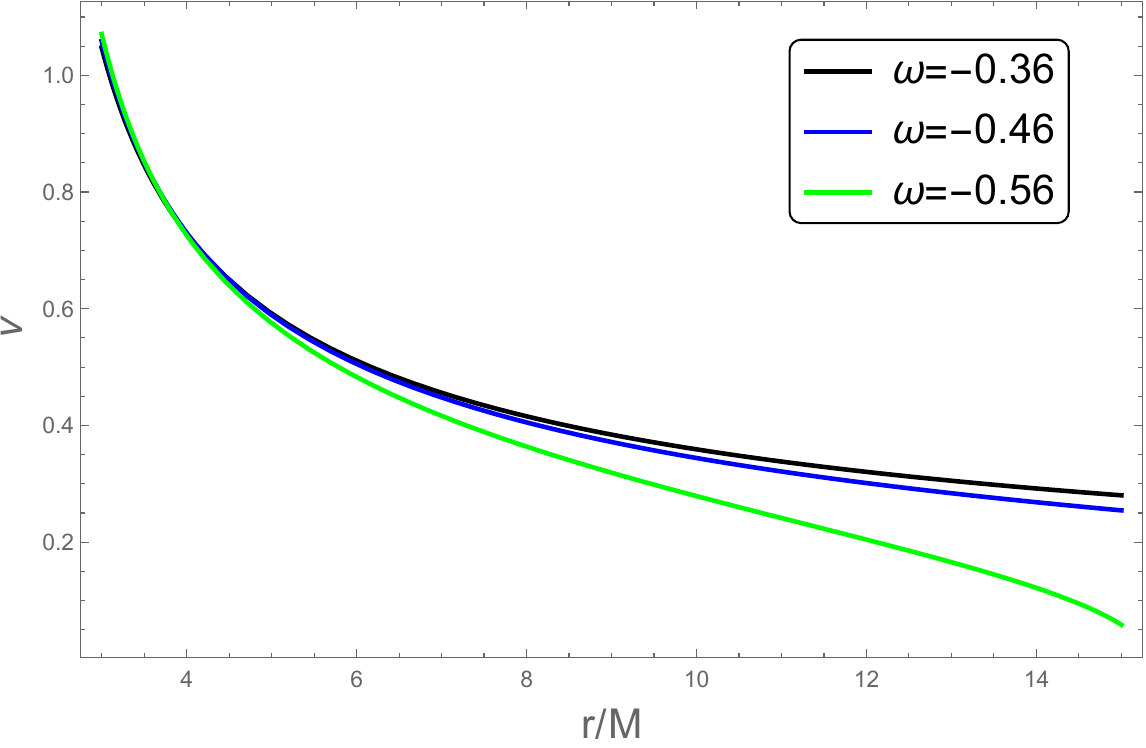}
\caption{The orbital velocity $v$ is plotted along $r/M$ for $M=1$: 
a. $Q/M=1$, $c=0.0009$, $\omega =-0.39$ and various values of  $\nu$; 
b. $\omega=-0.39$, $c=0.0009$, $\nu=1$ and several values of $Q/M$;
b. $\nu=0.1$, $c=0.03$, $Q/M=0.1$, and different values of $\omega$.}\label{fig:3}
\end{figure*}
\begin{figure}[t] 
\centering
\includegraphics[scale=0.45]{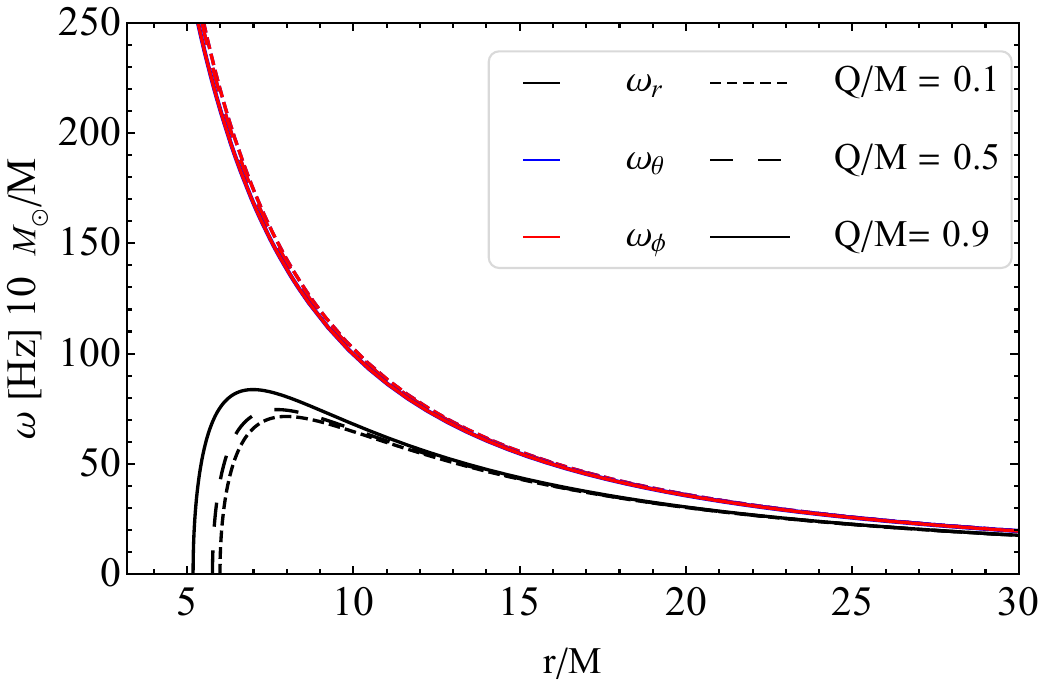}
\caption{\label{4}The profile of the fundamental frequencies is plotted against $r/M$ for neutral particles, shown for different values of the dyonic charge $Q/M$. }
\end{figure}
In Fig. \ref{fig:2}, we illustrate the radial profile of the Keplerian frequency for the motion of charged particles in a circular orbit around a weakly magnetized Einstein-ModMax BH with different parameter choices. In panel (a), it is shown that the Keplerian frequency increases near the BH as the parameter $\nu$ grows, following standard Keplerian scaling at larger distances from the BH. Panel (b) demonstrates that an increase in the parameter $\omega$ enhances the gravitational pull, leading to higher orbital motion near the BH and a rise in the Keplerian frequency $\Omega_{k}$ in the strong gravitational regime, which converges to Keplerian scaling at larger radii. Panel (c) highlights the effect of the dimensionless dyonic charge $Q/M$ on the Keplerian frequency $\Omega_{k}$, indicating that an increase in the dimensionless dyonic charge $Q/M$ results in a suppression of $\Omega_{k}$ near the BH due to electromagnetic repulsion counteracting the gravitational pull. While the asymptotic behavior of $\Omega_{k}$ is universal, variations in the parameters exhibit distinct effects in the vicinity of the BH.

Figure~\ref{fig:3} shows the radial profile of the orbital velocity \(v\) for test particles on circular orbits around the considered BH. From Fig.~\ref{fig:3}(a), we observe that increasing the ModMax parameter \(\nu\) raises orbital velocity \(v\) in the strong-gravity region and this effect becoming more pronounced as the particle approaches the BH horizon; correspondingly, the curves shift upward for larger \(\nu\) and merge at large distance $r/M$. Figure~\ref {fig:3}(b) shows that, as the electromagnetic charge \(Q/M\) increases, the orbital velocity \(v\) near the BH horizon is suppressed. This occurs because electromagnetic charges contribute as a repulsive \(+Q^{2}/r^{2}\) term that weakens gravitational attraction. Additionally, we have noted the influence of the quintessence state parameter $\omega$ in Fig.~\ref{fig:3}(c). It is evident that an increase in the quintessence state parameter $\omega$ (from -0.56 to -0.36) results in a higher orbital velocity $v$ in the strong gravitational region because the quintessence parameter strengthens the gravitational attraction. We also examine the behavior of the Keplerian frequency $\Omega_{k}$ and the orbital velocity $v$ at the ISCO, with the corresponding values presented in Tables~\ref{tab:ISCO_nu_w_split} and \ref{tab:ISCOvalues}.

 \section{The frequencies of epicyclic motion} \label{Sec:qp0}
 
This section focuses on the epicyclic motion of test particles, arising from small perturbations in radial $r = r_{0} + \delta r $ and vertical $\theta = \frac{\pi}{2} + \delta \theta$ directions near the stable circular orbits in the considered BH spacetime. The particle oscillates with radial and latitudinal epicyclic frequencies, governed by linear harmonic oscillator equations that are described in terms of small perturbations $\delta r$ and $\delta \theta $ as 
\begin{eqnarray}
\ddot{\delta r} + \bar{\Omega}_{r}^{2}\,\delta r = 0 
\mbox{~~and~~}\ddot{\delta \theta} + \bar{\Omega}_{\theta}^{2}\,\delta \theta = 0, \label{22}
\end{eqnarray}
where $\bar{\Omega}_{r}$ is the radial frequency and $\bar{\Omega}_{\theta}$ represents the vertical frequency. These frequencies obtained from Eq. (\ref{4}), as \cite{Shaymatov:2020yte, Stuchlik:2021guq}
\begin{eqnarray}
\bar{\Omega}_{r}^{2} &=& \frac{1}{g_{rr}} \frac{\partial^{2} H_{\text{pot}}}{\partial r^{2}}, \label{23}\\
\bar{\Omega}_{\theta}^{2} &=& \frac{1}{g_{\theta \theta}} \frac{\partial^{2} H_{\text{pot}}}{\partial \theta^{2}}, \label{24}\\
\bar{\Omega}_{\phi} &=& \frac{1}{g_{\phi \phi}} \left( \mathcal{L} - \frac{q}{m} A_{\phi} \right). \label{25}
\end{eqnarray}
As previously mentioned, the periodic motion is possible when the particles move in a stable circular orbit around the BH, characterized by fundamental frequencies, specific energy, and specific angular momentum. In order to describe such motion, we consider $u^{\alpha}=(u^t, 0, 0, u^{\phi})$ along with normalization condition $u_{\alpha}^{\alpha}=-1$ from which we obtain the following relation 
\begin{eqnarray}
u^{t} &=& \frac{1}{\sqrt{-g_{tt} - \Omega^{2} g_{\phi\phi}}}\, ,\\ \label{26}
\mathcal{E} &=& -\frac{g_{tt}}{\sqrt{-g_{tt} - \Omega^{2} g_{\phi\phi}}} + \frac{g_{e}e^{-\nu}}{r}\, ,\\ \label{27}
\mathcal{L} &=& \frac{g_{\phi\phi} \, \Omega}{\sqrt{-g_{tt} - \Omega^{2} g_{\phi\phi}}} - \sigma_{m}(1-\cos\theta), \label{28}
\end{eqnarray}
where $\Omega=\frac{d\phi}{dt}$ corresponds to the Keplerian frequency obtained from Eq. (\ref{18}). Moreover, by using Eqs. (\ref{23}-\ref{28}) along with Eq. (\ref{18}) we determine the $\Omega_{\theta}$ and $\Omega_{r}$ expressed as
\begin{eqnarray}
  \Omega^{2}_{\theta}&&=\Big[c e^{\nu } r(r^2 (3 \omega +1)-3 \sigma_{m} ^2 (\omega +1))+2 r^{3 \omega }(e^{\nu } r(M r^2+\nonumber\\&&\sigma_{m} ^2 (r-3 M))-Q^2(r^2-2 \sigma_{m} ^2))\Big]\Big[2 r^{3 \omega +4}(e^{\nu } r (r-3 M)\nonumber\\&&+2 Q^2)-3 c e^{\nu } r^5 (\omega +1)\Big]^{-1},\label{29}  
\end{eqnarray} 
and
\begin{eqnarray}
  \Omega^{2}_{r}&&=\Big[e^{-\nu } r^{-3 \omega -4}(-3 c^{2} e^{2 \nu } r^{2} (\omega +1) (3 \omega +1)\nonumber\\&&+c e^{\nu } r^{3 \omega +1}(e^{\nu }(6 g_{e} r \omega  \Big(\Big(6 r(-c (\omega +1) r^{-3 \omega }-2 M)\nonumber\\&&+8 e^{-\nu } Q^{2}\Big)r^{-2}+4)^{1/2}+3 g_{e}^{2} (\omega +1)+r(6 M (\omega  \nonumber\\&&(3 \omega -4)-2)-9 r \omega^{2}+r))+3 Q^2 (\omega  (8-3 \omega )\nonumber\\&&+3))+2 r^{6 \omega }(-e^{\nu } Q^{2}(g_{e} r \Big(\Big(-6 c (\omega +1) r^{1-3 \omega }-\nonumber\\&&12 M r+8 e^{-\nu } Q^{2}\Big)r^{-2}+4\Big)^{1/2}+2 g_{e}^{2}-9 M r)+\nonumber\\&&e^{2 \nu } r(g_{e} r^{2} \Big(\Big(-6 c (\omega +1) r^{1-3 \omega }-12 M r+8 e^{-\nu }\nonumber\\&& Q^{2}\Big)r^{-2}+4\Big)^{1/2}+g_{e}^{2} (3 M-r)+M r (r-\nonumber\\&&6 M))-4 Q^{4}))\Big]\Big[2 r^{3 \omega } (e^{\nu } r (r-3 M)+2 Q^{2})\nonumber\\&&-3 c e^{\nu } r (\omega +1)\Big]^{-1}.\label{30}  
\end{eqnarray}
\begin{figure*}
\includegraphics[scale= 0.3]{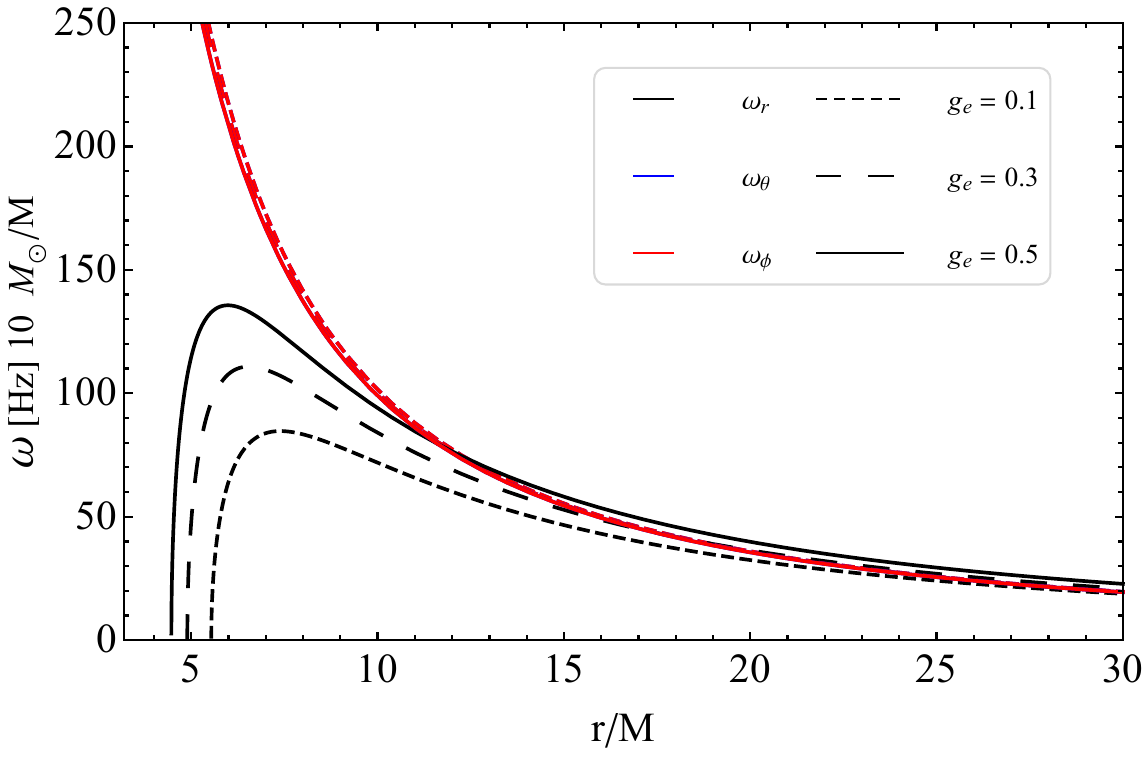}   
\includegraphics[scale= 0.3]{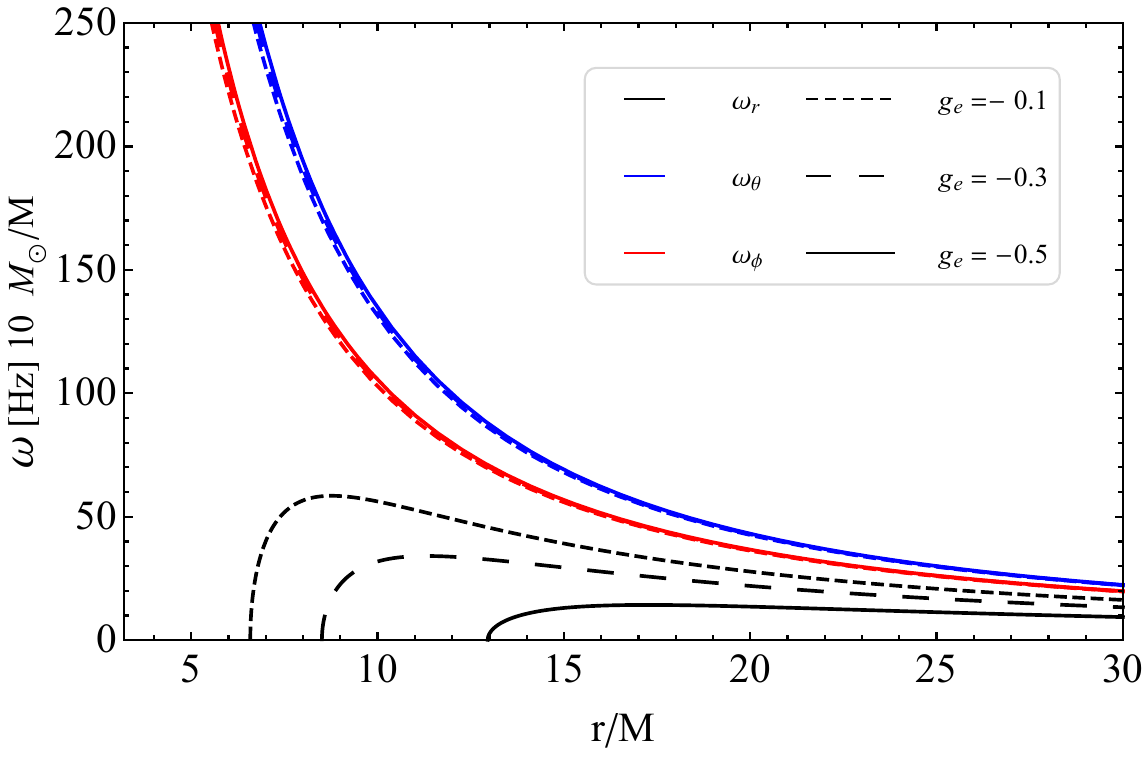}
\includegraphics[scale= 0.3]{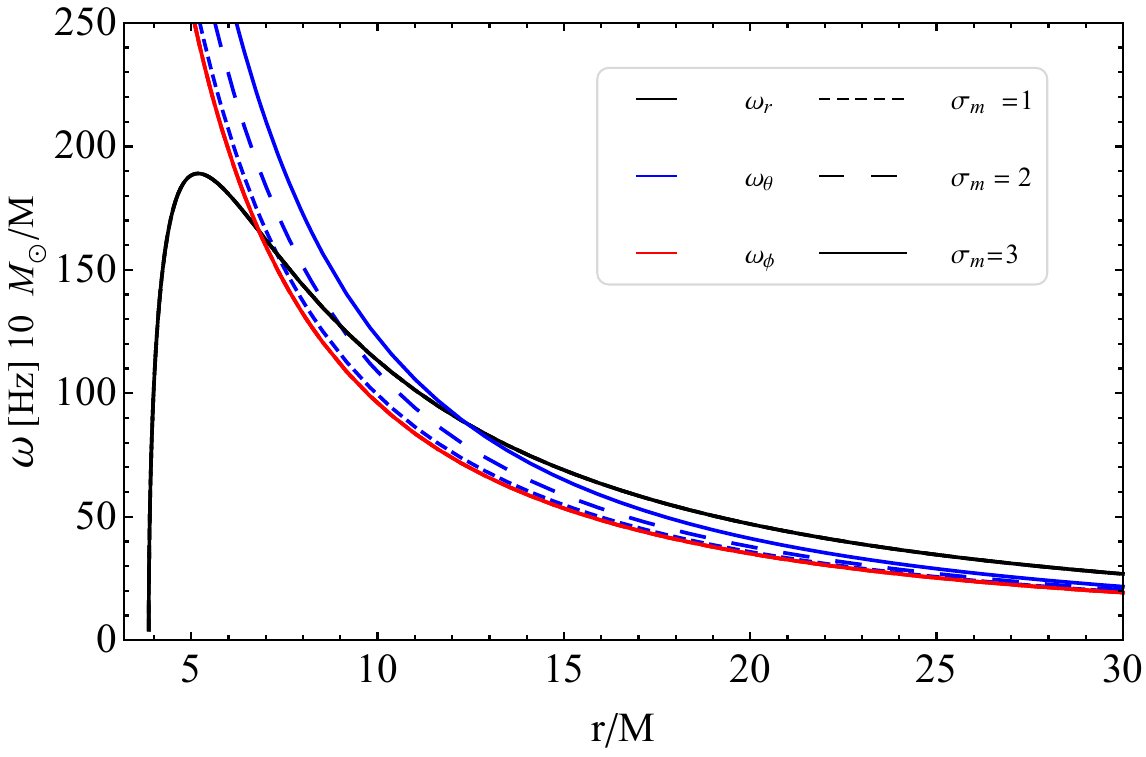}
\caption{The behavior of the fundamental frequencies as functions of $r/M$ for the case of charged particles is plotted. The left and middle panels are plotted for $M = 1$, $Q = 0.1$, quintessence state parameter $\omega = -0.4$, $\nu = 0.5$, and $\sigma_{m} = 3$, with various positive and negative values of the coupling charge $g_{e}$, while the right panel is plotted for $g_{e} = 1.0$ and different values of the magnetic coupling parameter $\sigma_{m}$.}\label{5}
\end{figure*}

\begin{figure*}
\centering
\includegraphics[scale= 0.3]{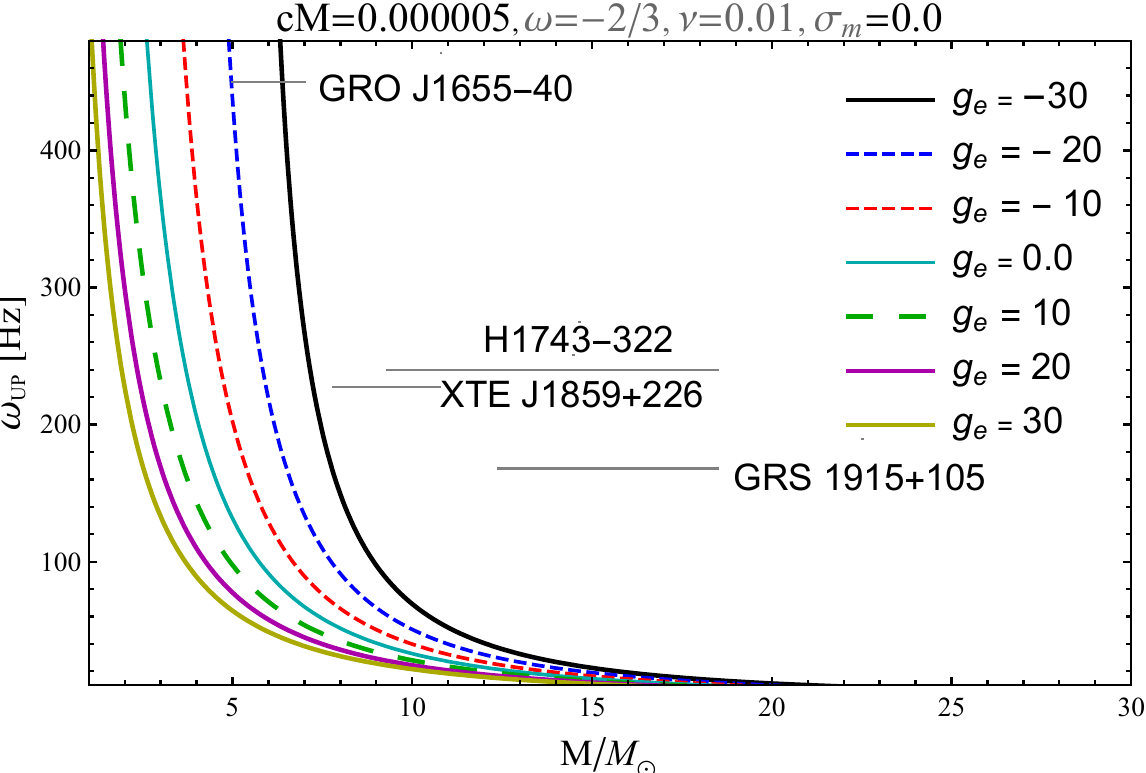}
\includegraphics[scale= 0.3]{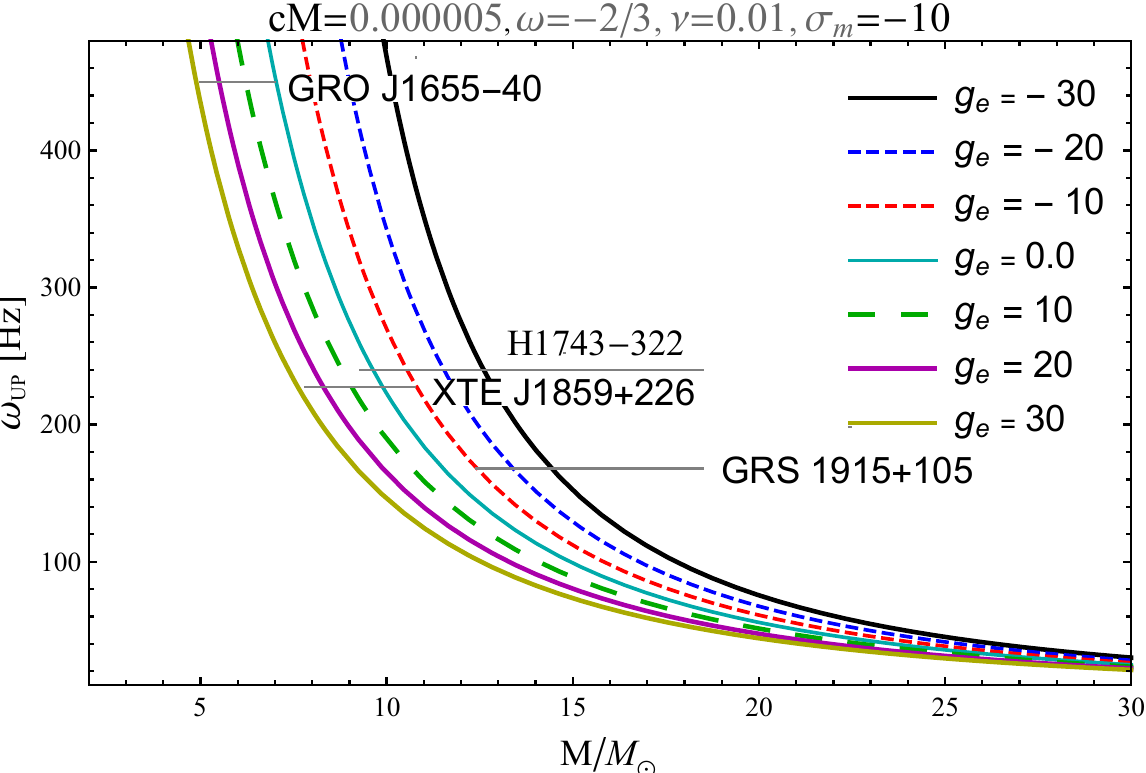}
\includegraphics[scale= 0.3]{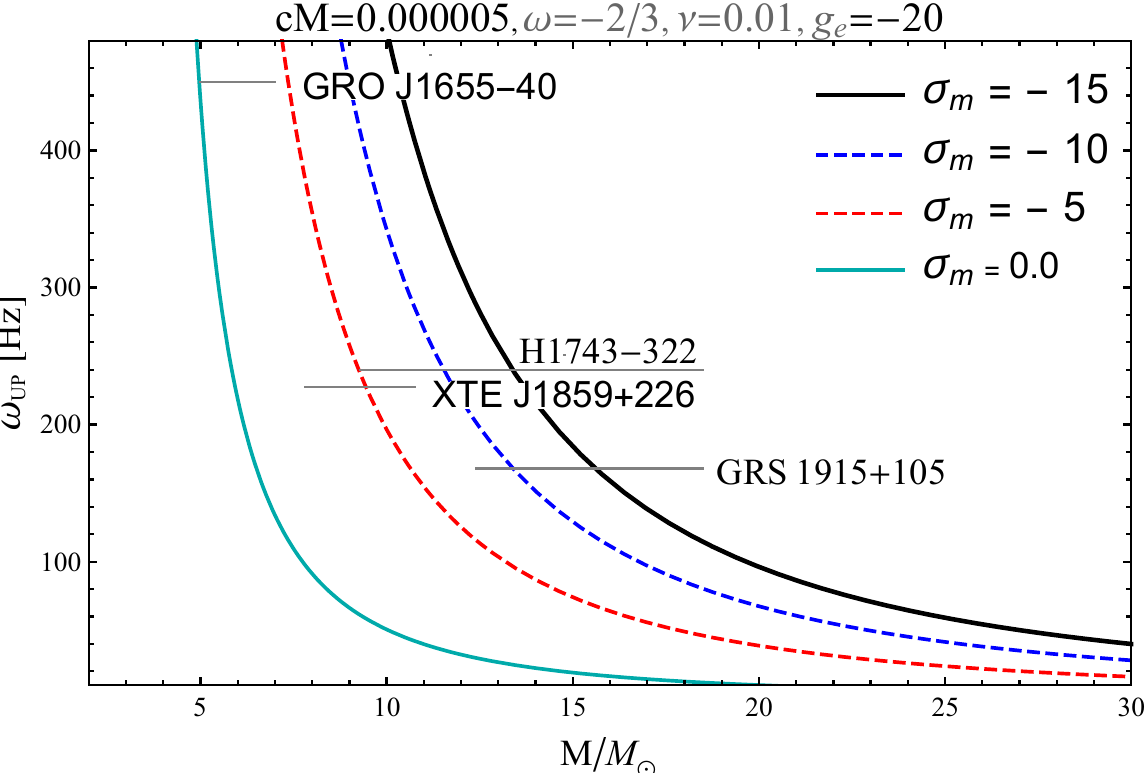}
\caption{
The behavior of the upper epicyclic frequency $\omega_{\mathrm{UP}}$ as a function of the mass ratio $M/M_{\odot}$ for various values of the coupling parameters $g_{e}$ and $\sigma_{m}$. In the left panel, $\omega_{\mathrm{UP}}$ is shown for $Q=0.01$ with $\sigma_{m}=0$ and several values of $g_{e}$. The middle panel presents the same quantity for $\sigma_{m}=-10$ and varying $g_{e}$. The right panel displays the dependence of $\omega_{\mathrm{UP}}$ for fixed $g_{e}=-20$ and different values of $\sigma_{m}$.
}
\label{6}
\end{figure*}
Our aim is to analyze the epicyclic frequencies mentioned above. As discussed earlier, the fundamental frequencies can be measured by a local observer, but they must also be measured by a distant observer located far from the BH vicinity at spatial infinity. To achieve this, we introduce a redshift factor and transform the frequencies from $\bar{\Omega}_i$ to $\omega_i$ for measurement at spatial infinity. This transformation is necessary for distant observers to accurately measure the frequencies. The relation between the two frequencies can be expressed using the redshift factor and constants $G$ and $c$ \cite{Shaymatov:2020yte, Shaymatov:2022enf, Stuchlik:2021wdh}. 
\begin{eqnarray}
\mathcal{\omega} = \frac{1}{2\pi}\,\frac{c^{3}}{GM}\,\frac{\bar{\Omega}}{(-g^{tt})\,\mathcal{E}}\,.
\end{eqnarray}

Thus, the observer situated at spatial infinity can directly measure these fundamental epicyclic frequencies and examine their behavior by considering QPOs as the primary source, which provides information about the accretion disk surrounding the supermassive BH.
Now we are interested in investigating the behavior of the epicyclic frequencies as measured by an observer located at spatial infinity. The radial profiles of the fundamental epicyclic frequencies for neutral and electromagnetically charged particles moving in stable circular orbits in the vicinity of the weakly magnetized Einstein–Maxwell BH are shown in Figs. \ref{4} and \ref{5}. Figure~\ref{4} illustrates the influence of the dyonic charge $Q = Q_{e} + Q_{m}$ on the fundamental epicyclic frequencies for neutral test particles. As shown in the figure, an increase in the dyonic charge $Q/M$ leads to a monotonic rise in the radial epicyclic frequency $\omega_{r}$ measured by an observer at spatial infinity, while both the orbital frequency $\omega_{\phi}$ and the vertical epicyclic frequency $\omega_{\theta}$ exhibit a decreasing trend with increasing $Q/M$. From Fig.~\ref{5}, we observe that the fundamental epicyclic frequencies depend sensitively on the charge–coupling parameters. From the left and middle panels of Fig.~\ref{5}, it is clear that as the values of the coupling parameter $g_{e}$ increase, for both positive and negative choices of $g_{e}$, the radial
epicyclic frequency $\omega_{r}$ increases, while the vertical frequency $\omega_{\theta}$ and the orbital angular frequency $\omega_{\phi}$ both decrease. This behavior demonstrates that the electromagnetic coupling $g_{e}$ modifies the effective potential in such a way that it enhances radial oscillations while suppressing the vertical and azimuthal oscillatory modes. Additionally, the right panel of Fig.~\ref{5} shows that, as the value of the coupling parameter $\sigma_{m}$ increases, the radial epicyclic frequency $\omega_{r}$ decreases, whereas both the vertical frequency $\omega_{\theta}$ and the orbital angular frequency $\omega_{\phi}$ increase. This trend indicates that the coupling $\sigma_{m}$ has an opposite effect compared to $g_{e}$ on the dynamical behavior of circular orbits.

Also, it is observed that quasi-periodic oscillations provide a powerful tool for testing theoretical models and probing unknown aspects associated with precise measurements of BH parameters. In this context, the X-ray power detected in microquasars, known as high-frequency quasi-periodic oscillations (HF QPOs), is considered to arise from the inner regions of low-mass X-ray binary systems. These systems, consisting of either a neutron star or a BH, play an increasingly significant role in astrophysics. In the strong-field regime near the ISCO radius, particles undergo radial and vertical oscillations with characteristic frequencies, giving rise to the quasi-periodic power spectra observed by a distant observer. Also, we have plotted the mass-frequency relation for the upper frequency of the QPO model with respect to the observational data for various values of the coupling parameters $g_{e}$ and $\sigma_{m}$, as shown in Fig.~\ref{6}. In the left panel, we set $\sigma_{m}=0$ and vary the coupling charge $g_{e}$. It is observed that as $g_{e}$ increases, the upper frequency decreases, and the contribution of $g_{e}$ alone is insufficient to reproduce the observed data. Moreover, the middle and right panels of Fig.~\ref{6} demonstrate that the combined effect of both coupling parameters $g_{e}$, and $\sigma_{m}$ becomes increasingly significant in achieving a best fit to the observational constraints. In particular, the presence of a nonzero magnetic coupling $\sigma_{m}$ shifts the upper-frequency profile toward smaller radii in order to best fit the observational data. Hence, the best fits arise from the simultaneous contribution of both coupling parameters, especially when $\sigma_{m}$ is increased.

\section{Parameter constraints for weakly magnetized BH surrounded by quintessence in Einstein–ModMax theory}~\label{Sec:parameter}

In this work, we consider the forced resonance mode \cite{Lee:2004bp, KluzniakAbramowicz2001, abramowicz2001precise, Banerjee:2022ffu} in which the twin high-frequency QPOs are interpreted as the result of nonlinear coupling between the vertical and radial epicyclic oscillations of matter in the accretion disk. When particles move near stable circular orbits, they undergo small oscillations in the vertical and radial directions with fundamental frequencies $\nu_{\theta}$ and $\nu_{r}$, respectively. These two modes interact nonlinearly and generate a combination (sum) oscillation that gives rise to the upper frequency, 
$\omega_{U} = \omega_{r} +  \omega_{\theta}$, while the lower frequency corresponds to the vertical epicyclic oscillation, $\omega_{L} = \omega_{\theta}$.
Moreover, we focus on the QPO observations of the X-ray binaries GRO~J1655--40, XTE~J1550--564, XTE~J1859+226, GRS~1915+105, and H1743--322, as listed in Table~\ref{tab: Ia}, to constrain the parameters of the weakly magnetized BH in the Einstein-ModMax theory. To determine the admissible range of BH parameters, we employ a Markov Chain Monte Carlo (MCMC) analysis, through which we obtain the best-fit values and the physically plausible parameter intervals.
\begin{table*}
\caption{The QPOs from the X-ray binaries that have been selected for investigation included their mass, orbital frequencies, periastron precession frequencies, and nodal precession frequencies.}
\label{tab: Ia}
\begin{ruledtabular}
\begin{tabular}{cccccc}
\; & GRO J1655-40   & XTE J1859+226 & GRS 1915+105  & H1743-322 
\\  
\hline 
$M~(M_{\odot})$ & $5.4 \pm 0.3$~\cite{Motta:2013wga} & $7.85 \pm 0.46$~\cite{Motta:2022rku} & $12.4^{+2.0}_{-1.8}$~\cite{Remillard:2006fc} & $\gtrsim 9.29$~\cite{Ingram:2014ara} 
\\ \;\\
$\nu_{\phi}$ (Hz) &$441 \pm 2$~\cite{Motta:2013wga}  & $227.5^{+2.1}_{-2.4}$~\cite{Motta:2022rku} & $168 \pm 3$~\cite{Remillard:2006fc} & $240 \pm 3$~\cite{Ingram:2014ara}
\\ \;\\                  
$\nu_{\text{per}}$ (Hz) & $298 \pm 4$~\cite{Motta:2013wga}  & $128.6^{+1.6}_{-1.8}$~\cite{Motta:2022rku} & $113 \pm 5$~\cite{Remillard:2006fc} & $165^{+9}_{-5}$~\cite{Ingram:2014ara}  
\\\;\\
$\nu_{\text{nod}}$ (Hz) &  $17.3 \pm 0.1$~\cite{Motta:2013wga}  & $3.65 \pm 0.01$~\cite{Motta:2022rku}&-- & $9.44 \pm 0.02$~\cite{Ingram:2014ara}
\end{tabular}
\end{ruledtabular}
\end{table*}
\begin{figure*}
\centering
\includegraphics[scale= 0.2]{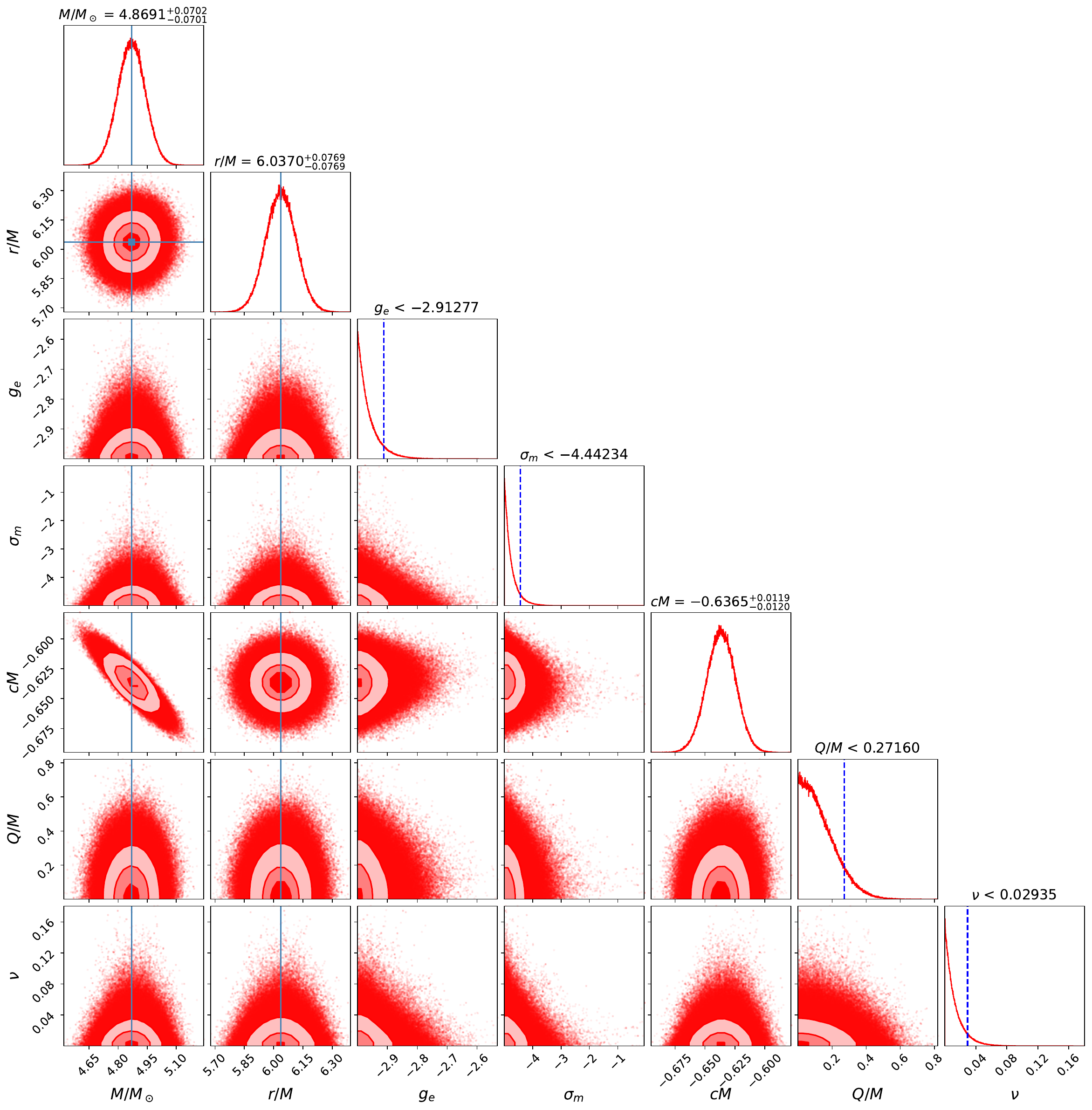}
\includegraphics[scale= 0.2]{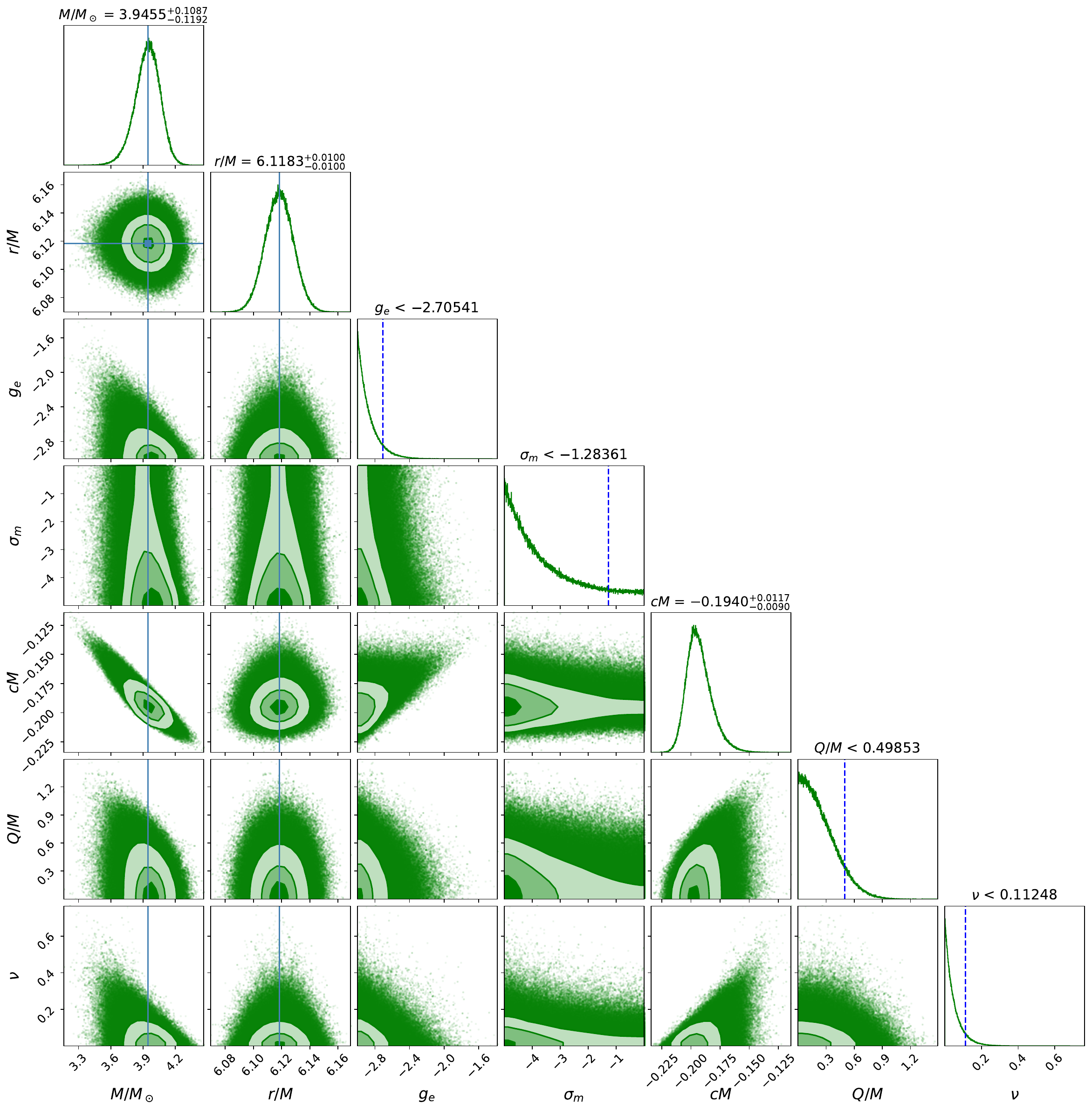}
\includegraphics[scale= 0.2]{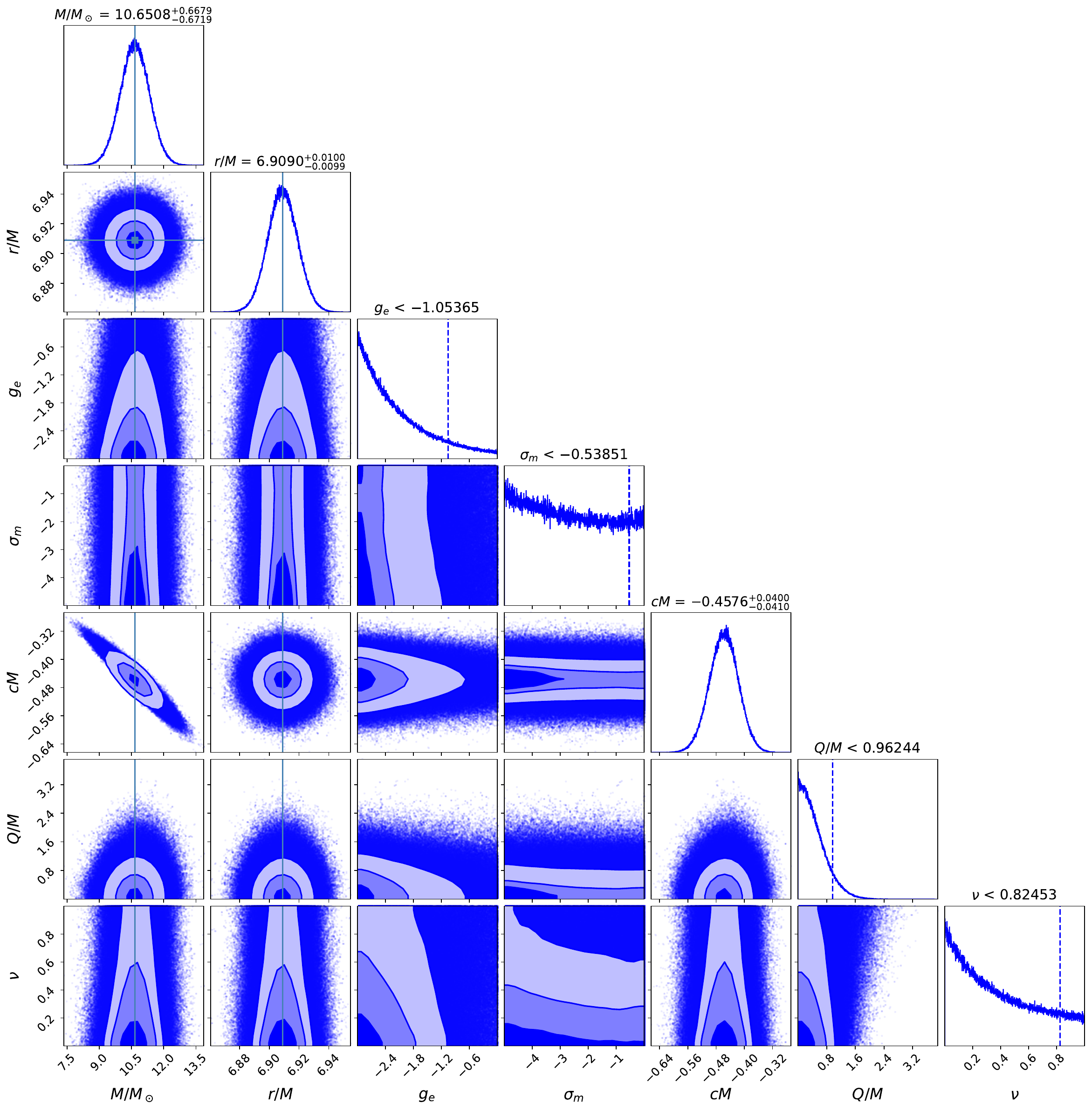}
\includegraphics[scale= 0.2]{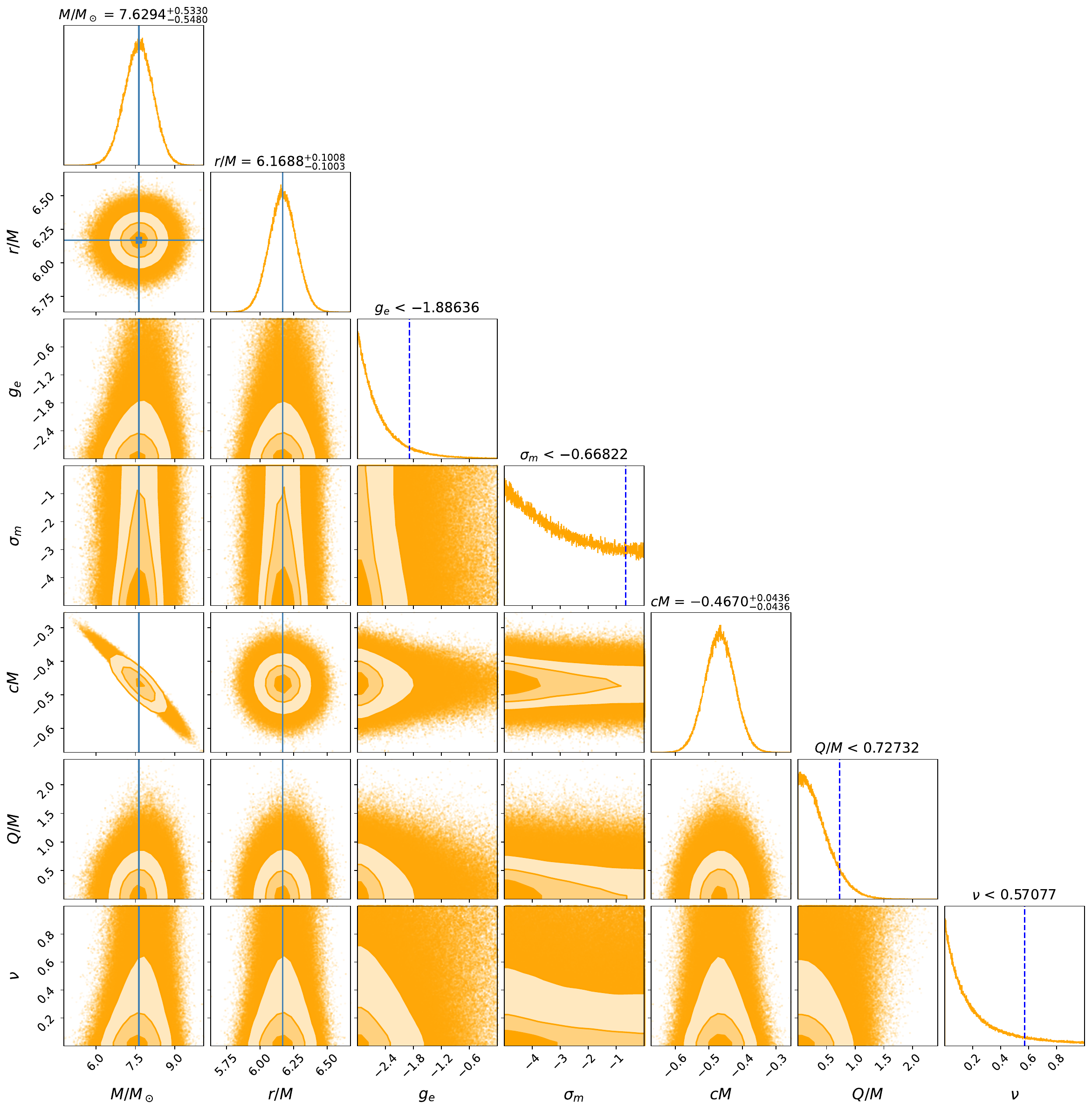}
\caption{Marginalized posterior distributions and two-dimensional joint confidence regions for the parameters of the considered BH, obtained from the MCMC analysis of QPO data for GRO~J1655--40 (red contours), XTE~J1859+226 (green contours), GRS~1915+105 (blue contours), and H1743--322 (orange contours). The BH mass $M$, orbital radius $r/M$, and the quintessence parameter are constrained at the $68\%$ confidence level, while the remaining parameters are constrained at the $90\%$ confidence level.
}\label{fig: 6}
\end{figure*}
\begin{table*}
 \caption{The best-fit values of the considered BH parameters $M$ and $r/M$ are reported with 68\% confidence intervals based on QPO data from X-ray binaries, while the parameters $g_{e}$, $\sigma_{m}$, $cM$, $Q/M$, and $\nu$ are constrained at the 90\% confidence level.}
\label{tab: VI}
\begin{ruledtabular}
\begin{tabular}{cccccc}\
\; & GRO J1655-40 \ & XTE J1859+226 \ & GRS 1915+105 \ & H1743-322
\\ \hline  \\
$M/(M_{\odot})$ & $4.8691^{+0.0702}_{-0.0701}$  & $3.9455^{+0.1087}_{-0.1192}$ & $10.6508^{+0.6670}_{-0.6719}$& $7.6294^{+0.5330}_{-0.5480}$ 
\\ \;  \\
 $r/M$ &$6.0370^{+0.0.0769}_{-0.0769}$ & $6.1183^{+0.0100}_{-0.0100}$ & $6.679090^{+0.0100}_{-0.0088}$ & $6.1688^{+0.10080}_{-0.1003}$ 
 \\ \; \\
 $g_{e}$ & $<-2.9127$ & $<-2.70541$ & $<-1.05363$ & $<-1.88636$
 \\\;\\
 $\sigma_{m}$ &  $<-4.44234$   &$<-1.28361$&  $< -0.53851$ & $<-0.66822$
 \\\;\\
$cM$ & $-0.6365^{+0.0117}_{-0.0120}$   & $-0.6365^{+0.0117}_{-0.0120}$  & $-0.4576^{+0.0400}_{-0.0410}$  & $-0.4670^{+0.0436}_{-0.0436}$ 
\\\;\\
 $Q/M$ & $<0.2716$    & $<0.49853$   & $<0.96244$   &  $<0.72732$
 \\\;\\
  $\nu$ & $<0.02935$   & $<0.11248$   & $<0.82453$   &  $<0.57077$
\end{tabular}
 \end{ruledtabular}
\end{table*}

\begin{figure}
\centering
\includegraphics[scale=0.5]{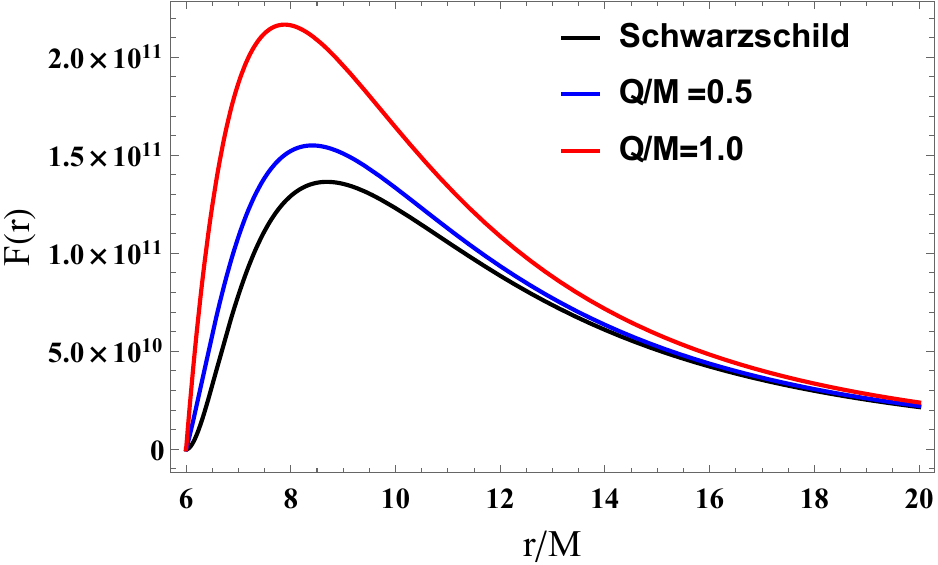}
\includegraphics[scale=0.5]{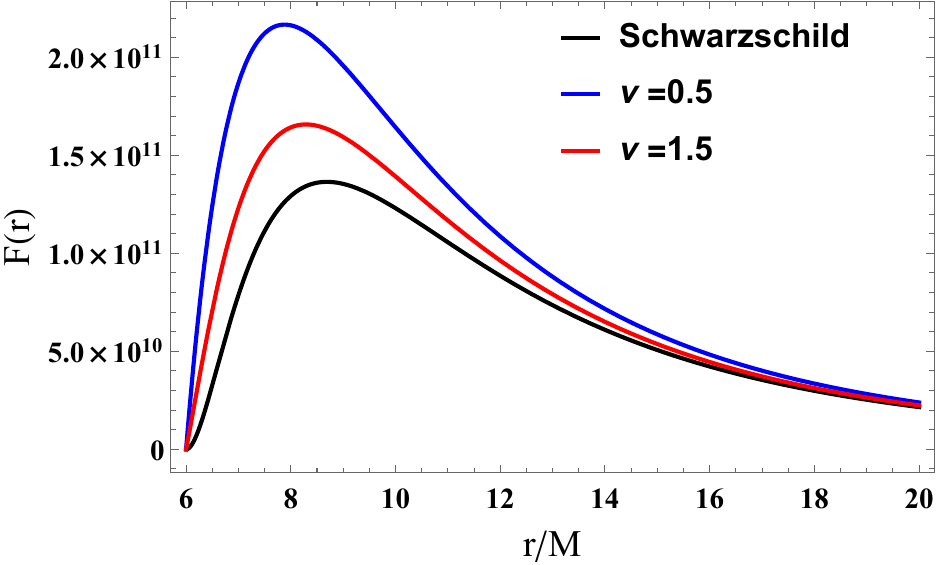}
\includegraphics[scale=0.5]{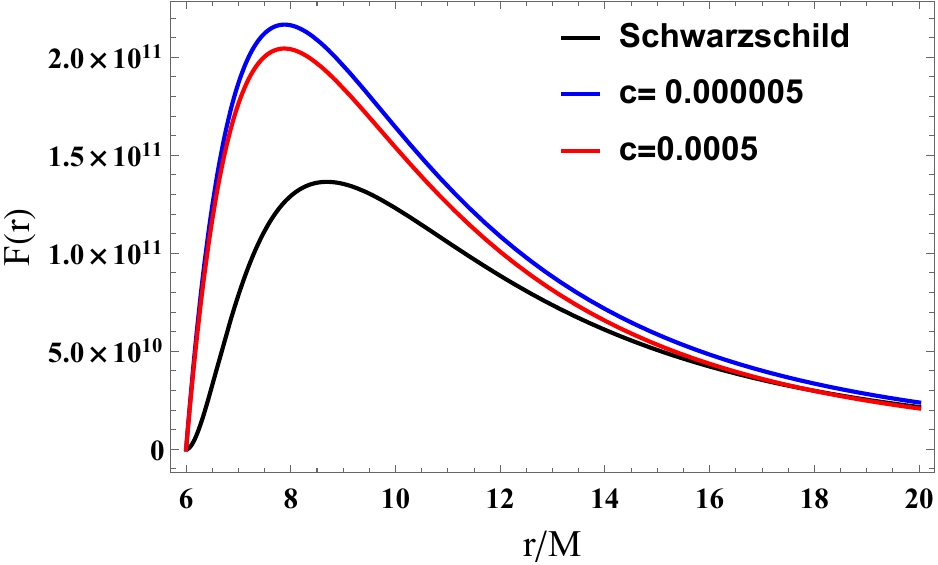}
\caption{Radial profiles of the energy flux $\mathcal{F}(r)$ emitted from the accretion disk around a weakly magnetized BH in Einstein-ModMax theory surrounded by quintessence for $M=1$ (mass), $\omega=-2/3$ (quintessence state parameter), $\sigma_{m}=0.1$ (magnetic coupling parameter), $g_{e}=0.1$ (electric coupling parameter). The upper panel shows the effect of the dyonic charge $Q/M$ by setting $\nu=0.5$ and $c=0.000005$, the middle panel indicates the impact of the ModMax parameter $\nu$, by setting $Q=1$, $c=0.000005$, and the bottom panel represents the influence of the quintessence parameter $c$ by setting $\nu=0.5$ and $Q=1$.}
\label{fig:7}
\end{figure}

\subsection{Markov chain Monte Carlo analysis}
This subsection focuses on the MCMC analysis implemented in the \emph{emcee} package~\cite{Foreman-Mackey:2012any} to constrain parameters of the weakly magnetized BH in the Einstein-ModMax theory. The posterior distribution is provided by
\begin{eqnarray}
\mathcal{P}(\Theta | \mathcal{D}, \mathcal{M}) = \frac{P(\mathcal{D} | \Theta, \mathcal{M}) \, \pi(\Theta | \mathcal{M})}{P(\mathcal{D} | \mathcal{M})},
\end{eqnarray}
Here, $\pi(\Theta \mid \mathcal{M})$ denotes the prior, $P(\mathcal{D} \mid \Theta, \mathcal{M})$ represents the likelihood, and $\mathcal{D}$ denotes the data vector, while $\mathcal{M}$ corresponds to the underlying model. For the priors of the parameters $(M, r/M)$, we use the truncated Gaussian distributions has the form
\begin{eqnarray}
\pi(\theta_i) \propto \exp\left[ -\frac{1}{2} \left( \frac{\theta_i - \theta_{0, i}}{\sigma_i} \right)^2 \right], \quad \theta_{\text{low}, i} < \theta_i < \theta_{\text{high}, i}, \nonumber \\
\end{eqnarray}
where $\theta_{i} = [M, r/M]$, and $\sigma_{i}$ denotes the standard deviation for the BH parameters. For the remaining parameters $(g_{e}, \sigma_{m}, c, q, \nu)$, we adopt uniform priors in our analysis. The total likelihood function $\mathcal{L}$ is given as \cite{Liu:2023vfh, Bambi:2013fea}
\begin{eqnarray}
\log \mathcal{L}_{\rm tot} = \log \mathcal{L}_{\rm U } + \log \mathcal{L}_{\rm L}
\end{eqnarray}
with
\begin{eqnarray}
\log \mathcal{L} = -\frac{1}{2} \sum_i \frac{\left(\Vec{D}_{\rm obs}^i - \Vec{D}_{\rm th}^i\right)^2}{(\sigma_i)^2}\, ,
\end{eqnarray}
where, $\Vec{D}_{\rm obs}^i$ is $i$-th observed, and $\Vec{D}_{\rm th}^i$ indicates the $i$-th theoretical data points, while $\sigma_i$ denotes the associated statistical measurement uncertainty. 

\subsection{Results and discussions}

In this subsection, we explore the seven-dimensional parameter space of a weakly magnetized BH surrounded by quintessence within the framework of Einstein-ModMax theory using a Markov Chain Monte Carlo (MCMC) analysis. Throughout this analysis, the quintessence equation-of-state parameter is fixed to $\omega=-2/3$. We consider QPO observations from the X-ray binary systems GRO J1655-40, XTE J1859+226, GRS 1915+105, and H1743-322. The best-fit values obtained from the MCMC analysis are summarized in Table~\ref{tab: VI}. Figure~\ref{fig: 6} presents the corresponding posterior distributions for all model parameters associated with the weakly magnetized BH surrounded by quintessence in Einstein--ModMax theory. In these plots, the BH mass $M$, the dimensionless orbital radius $r/M$, and the dimensionless quintessence parameter $c/M^{1+3\omega}$ are constrained at the $68\%$ confidence level, while the remaining parameters are reported at the $90\%$ confidence level. By applying an MCMC analysis, we obtained upper bounds at the $90\%$ confidence level on the ModMax electric coupling parameter $g_e$, the magnetic coupling parameter $\sigma_m$, the dyonic charge $Q$, and the nonlinear ModMax parameter $\nu$, as shown in Fig.~\ref{fig: 6}, with the corresponding upper bounds for the considered X-ray binaries presented in Table~\ref{tab: VI}. Our results imply that the observational data from the considered X-ray binaries do not require significant contributions from nonlinear electrodynamics, weak magnetization, or dyonic charge effects.

It is important to note that we obtained a nonzero signature for the quintessence parameter $c$ at the $68\%$ confidence level, which implies that the QPOs exhibit mild sensitivity to the surrounding dark-energy-like environment, as the quintessence field modifies the effective potential in a coherent manner over the relevant orbital region.

Overall, the results obtained from our analysis show that the QPO observations tightly constrain the fundamental black-hole parameters, namely the mass $M$ and the orbital radius $r/M$, while placing stringent upper limits on weak magnetization, nonlinear electromagnetic effects, and dyonic charge. This investigation reveals that the results are consistent with GR in the strong-field regime, while allowing only small deviations associated with Einstein-ModMax theory in the presence of a quintessence background.
\begin{figure}
\centering
\includegraphics[scale=0.5]{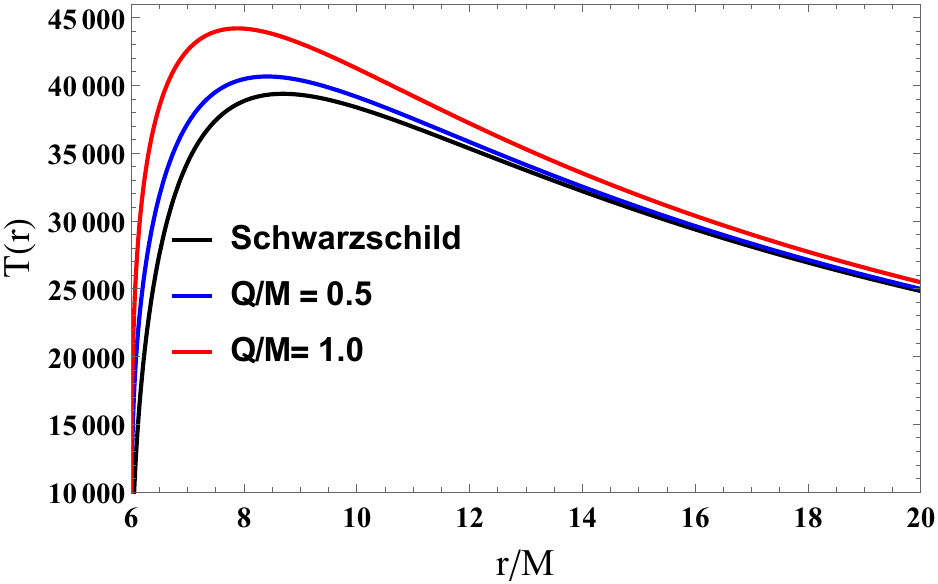}
\includegraphics[scale=0.5]{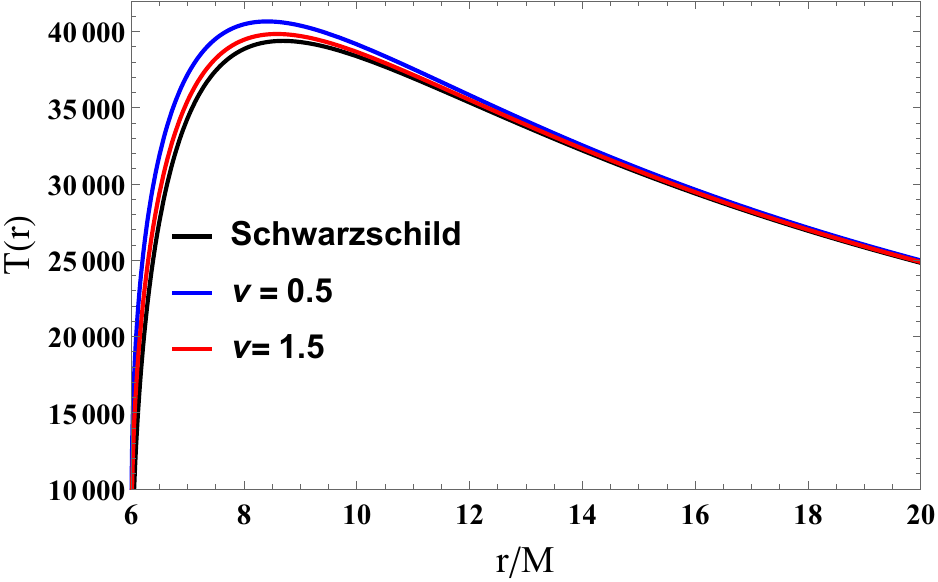}
\includegraphics[scale=0.5]{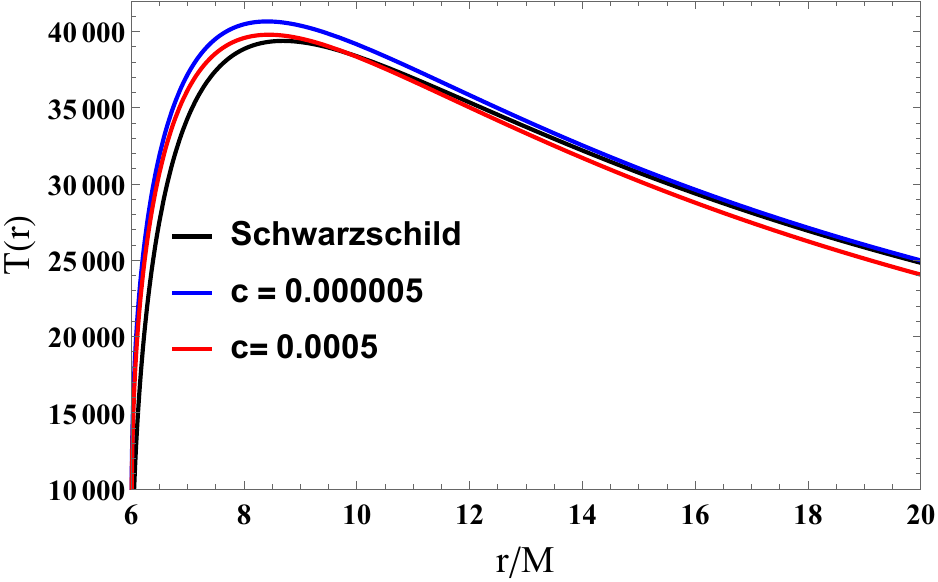}
\caption{The radial profiles of the temperature $T(r)$ of the accretion disk around a weakly magnetized BH in Einstein-ModMax theory surrounded by quintessence for $M=1$ (mass), $\omega=-2/3$ (quintessence state parameter), $\sigma_{m}=0.1$ (magnetic coupling parameter), $g_{e}=0.1$ (electric coupling parameter). Panel (a) shows the effect of the dyonic charge $Q/M$ by setting $\nu=0.5$ and $c=0.000005$, panel (b) indicates the impact of the ModMax parameter $\nu$, by setting $Q=1$, $c=0.000005$, and panel (c) represents the influence of the quintessence parameter $c$ by setting $\nu=0.5$ and $Q=1$.}
\label{fig:8}
\end{figure}

\begin{figure}
\centering
\includegraphics[scale=0.5]{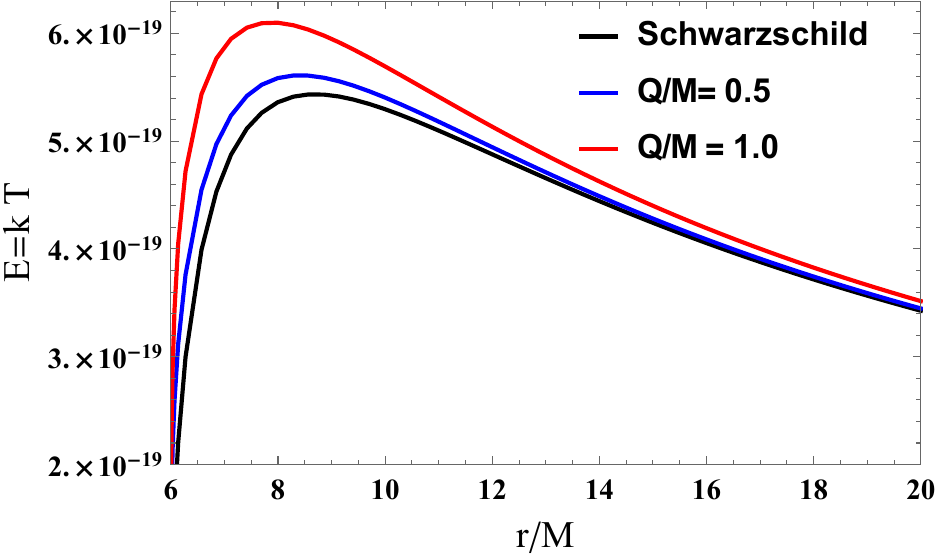}
\includegraphics[scale=0.5]{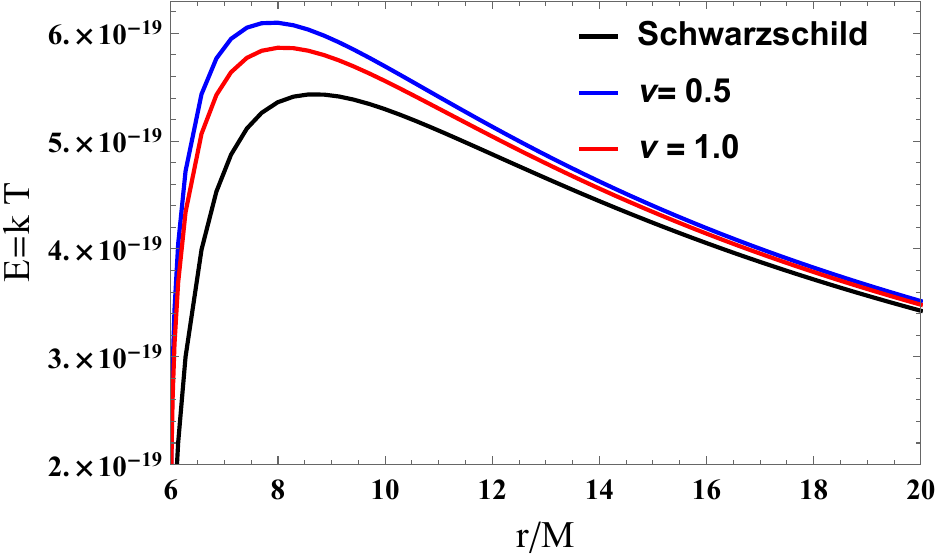}
\includegraphics[scale=0.5]{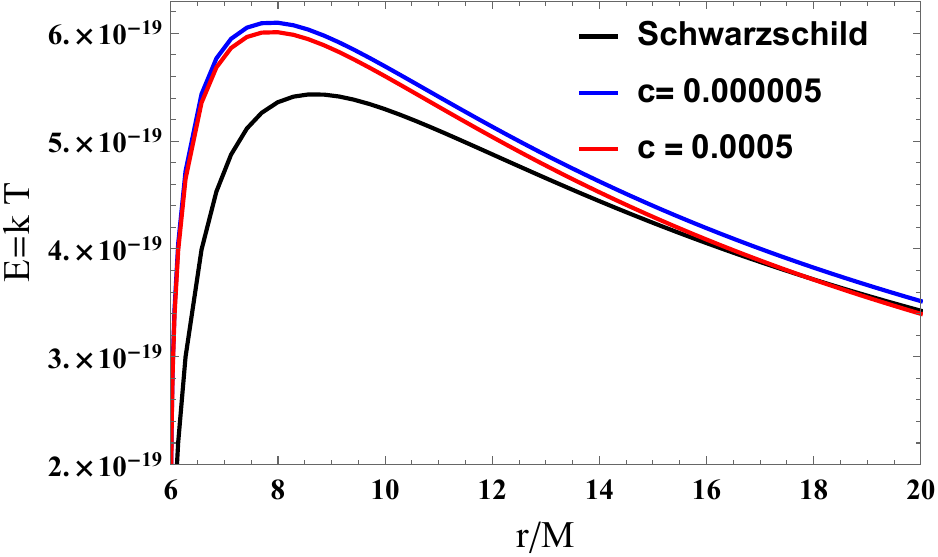}
\caption{The radial profiles of the energy $E=K\times T$ of the accretion disk around a weakly magnetized BH in Einstein-ModMax theory surrounded by quintessence for $M=1$ (mass), $\omega=-2/3$ (quintessence state parameter), $\sigma_{m}=0.1$ (magnetic coupling parameter), $g_{e}=0.1$ (electric coupling parameter). Panel (a) shows the effect of the dyonic charge $Q/M$ by setting $\nu=0.5$ and $c=0.000005$, panel (b) indicates the impact of the ModMax parameter $\nu$, by setting $Q=1$, $c=0.000005$,and panel (c) represents the influence of the quintessence parameter $c$ by setting $\nu=0.5$ and $Q=1$..}
\label{fig:9}
\end{figure}

\begin{figure*}
\label{fig:10}
\centering
\includegraphics[scale=0.5]{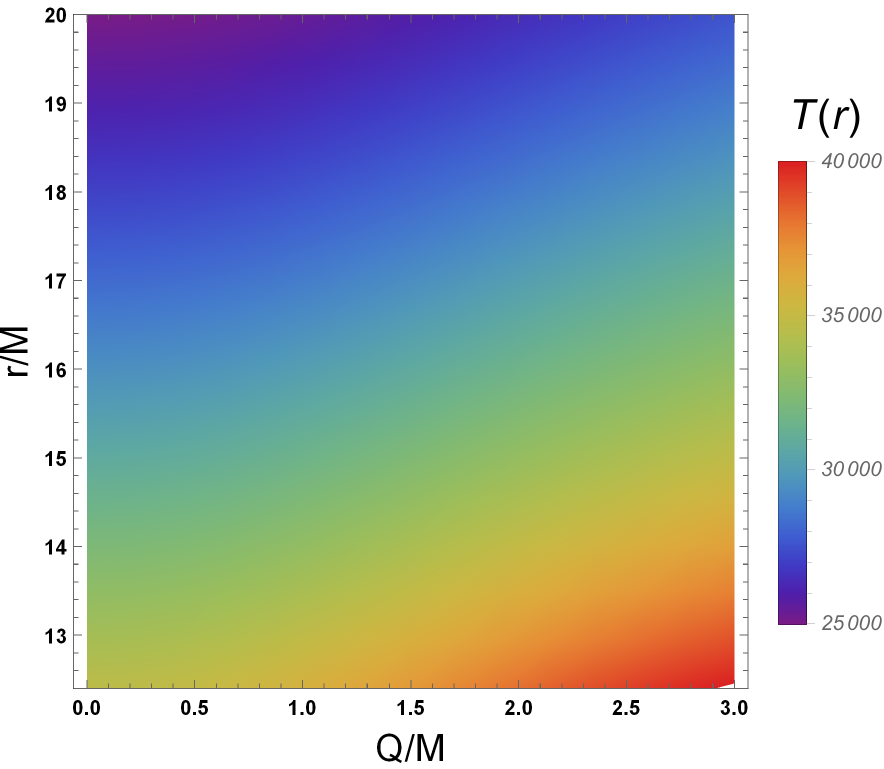}
\includegraphics[scale=0.5]{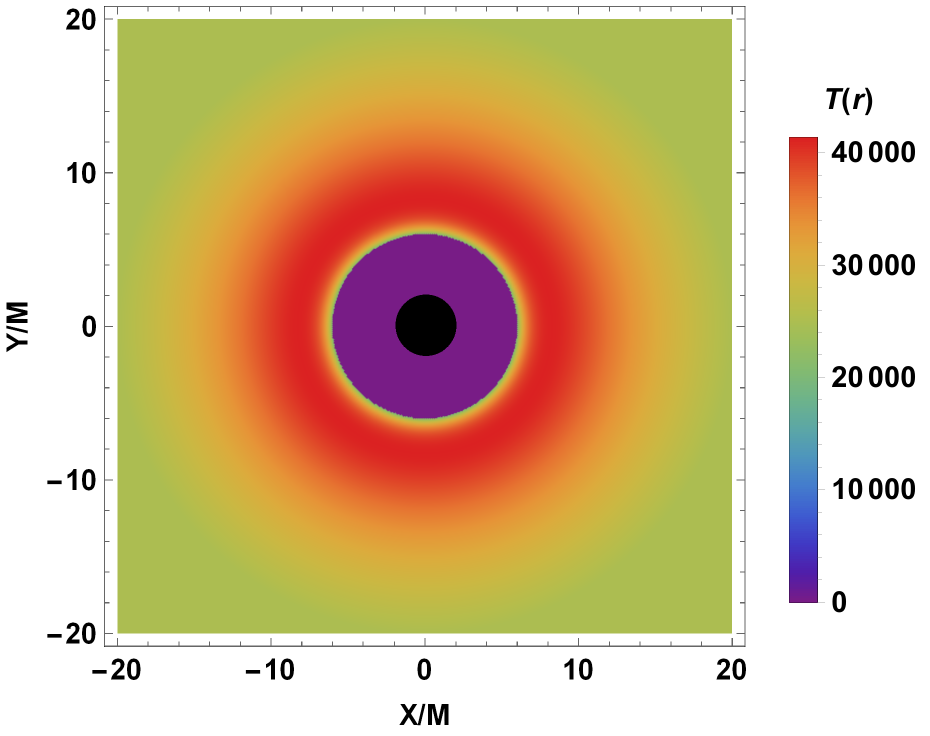}
\includegraphics[scale=0.5]{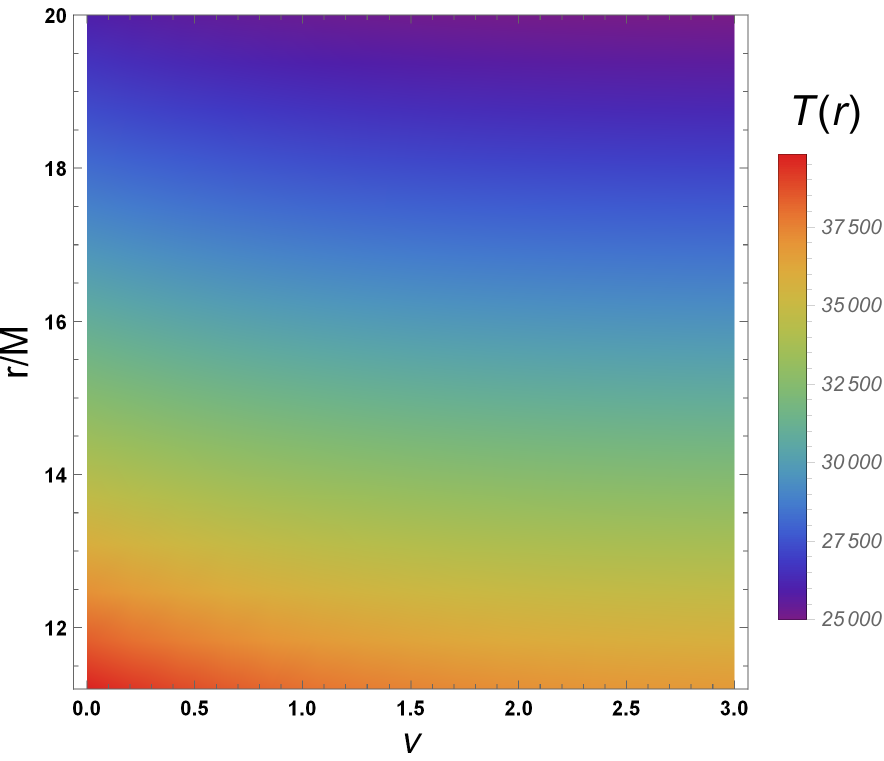}
\includegraphics[scale=0.5]{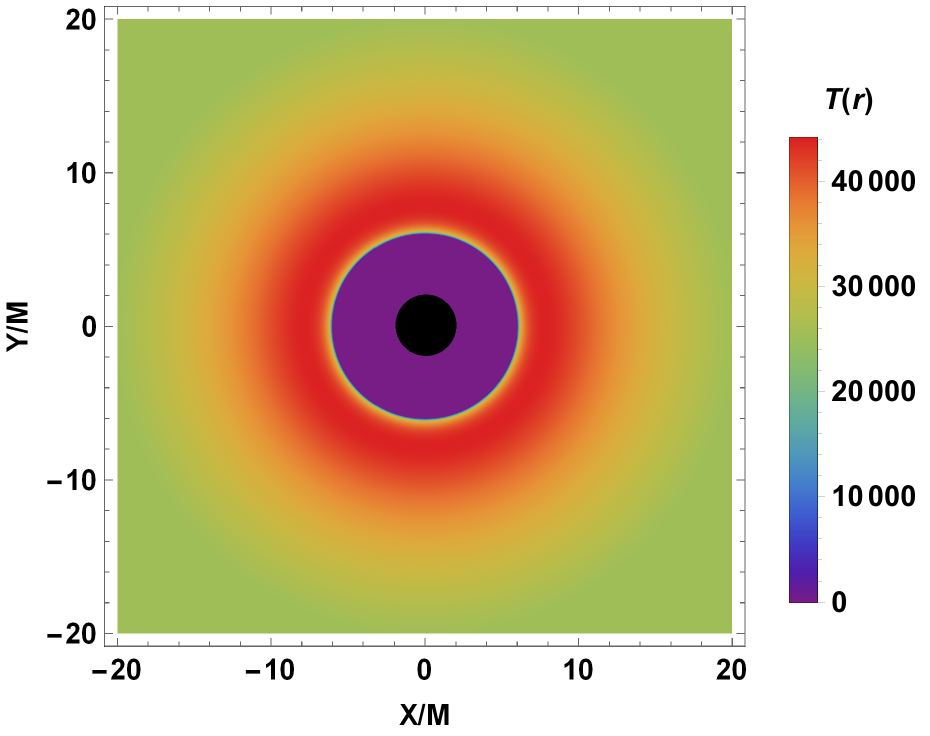}
\includegraphics[scale=0.5]{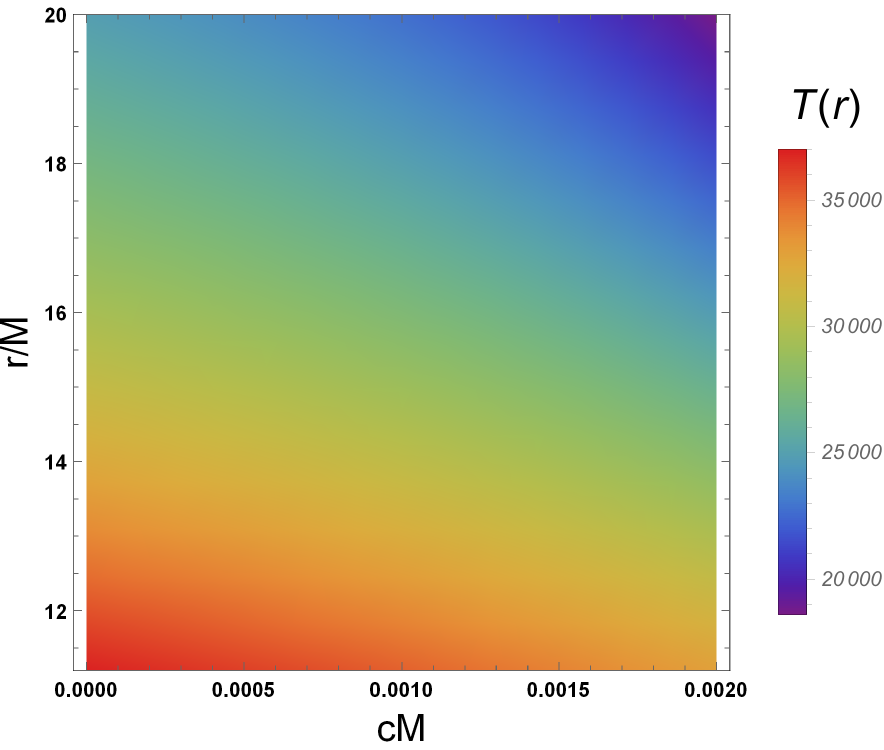}
\includegraphics[scale=0.5]{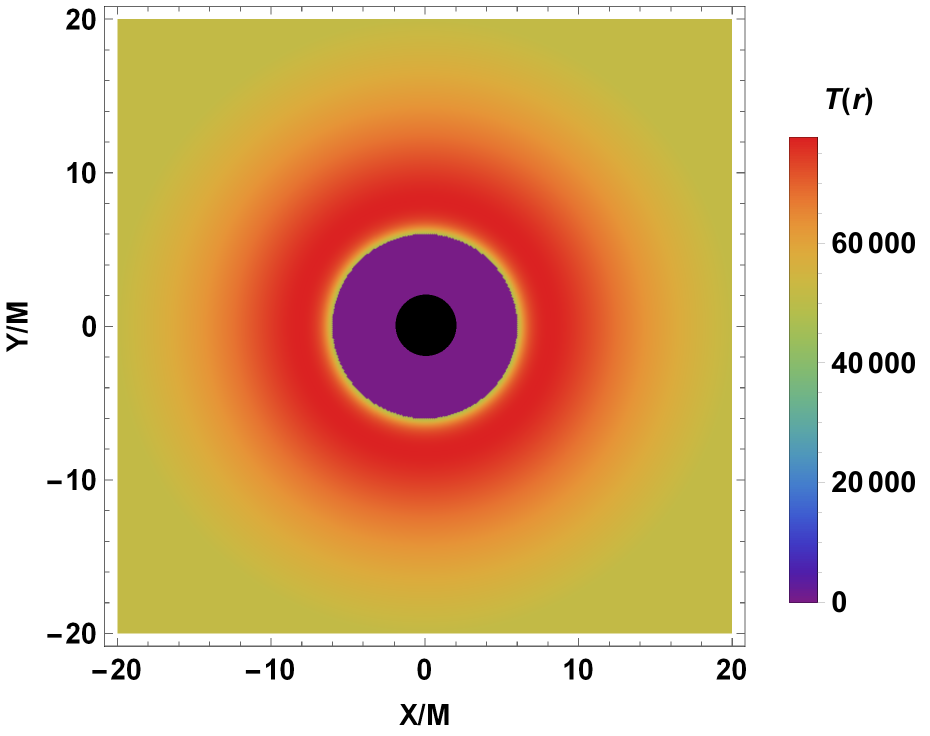}
\caption{The radial temperature profiles for the dimensionless dyonic charge $Q/M$, the ModMax parameter $\nu$, and the dimensionless quintessence parameter $cM$ are shown in the left panels. The corresponding temperature distributions on the equatorial $X$-$Y$ plane are displayed in the right panels in the form of density plots, where $X$ and $Y$ denote Cartesian coordinates.}
\label{fig:10}
\end{figure*}

\section{Accretion disk radiation for weakly magnetized BHs surrounded by quintessence within the framework of Einstein-ModMax theory}\label{Sec:Acc}

This section focuses on the flux produced by the accretion disk around a weakly magnetized BH surrounded by quintessence within the framework of Einstein-ModMax theory. The radiation emitted from the accretion disk of the considered BH is influenced by several physical effects. The ModMax parameter $\nu$ modifies the metric function through nonlinear electromagnetic contributions, while the weak magnetic field affects the motion of charged particles via electromagnetic coupling. In addition, the quintessence field modifies the gravitational potential, which influences the orbital structure of the accretion disk. The radiation of the accretion disk around the considered BH is computed by using the following expression \cite{Novikov:1973kta, Shakura:1972te, Thorne:1974ve, Boshkayev21PRD, Alloqulov24CPC}
\begin{eqnarray}
\mathcal{F}(r) = -\frac{\dot{M}_0}{4\pi\sqrt{\gamma}} \frac{\Omega_{,r}}{(E-\Omega L)^2} \int_{r_{\mathrm{ISCO}}}^{r} (E-\Omega L)\,L_{,r}\,dr, \nonumber \\
\end{eqnarray}
where $E$ denotes the specific energy, $L$ is the specific angular momentum, $\Omega$ is the angular velocity of the orbiting particles, $\sqrt{\gamma}=\sqrt{-g_{tt}\,g_{rr}\,g_{\phi\phi}}$, and $\dot{M}_0$ represents the mass accretion rate. For simplicity, we set $\dot{M}_0=1$. We have plotted the radial profile of the radiative flux emitted from the accretion disk of the considered BH in Fig.~\ref{fig:7}. From the figure, we observe that in all cases the flux rises from the ISCO radius, attains its maximum value in the strong-field region, and then declines at larger radii. The effects of the dyonic charge $Q/M$, the ModMax parameter $\nu$, and the quintessence parameter $c$ are shown in Fig.~\ref{fig:7}(a), Fig.~\ref{fig:7}(b), and Fig.~\ref{fig:7}(c), respectively. From the figure, it can be seen that the energy flux rises with increasing values of the charge $Q$. This shows that an increase in the $Q/M$ strengthens the gravitational potential, thereby enhancing the amount of gravitational binding energy converted into radiation. As a consequence, the energy flux emitted from the accretion disk rises. In Fig.~\ref{fig:7}(b) and Fig.~\ref{fig:7}(c), it is noted that the energy flux from the accretion disk decreases with increasing values of the ModMax parameter $\nu$ and the quintessence parameter $c$. This suppression occurs because both parameters $\nu$ and $c$ introduce repulsive contributions, which weaken the gravitational potential and consequently reduce the amount of gravitational binding energy converted into radiation.
Furthermore, the temperature of the accretion disk is determined from the relation $\mathcal{F}=\sigma T^{4}$, where $\sigma$ is the Stefan--Boltzmann constant. The disk temperature and the corresponding energy profiles are shown in Fig.~\ref{fig:8}. It is observed that the disk temperature $T$ and the energy $E = k \times T$ increase with increasing values of the charge $Q$, while they decrease with increasing values of both the ModMax parameter $\nu$ and the quintessence parameter $c$. For clarity, we present the temperature distribution using a color map in Fig.~\ref{fig:10}. From Fig.~~\ref{fig:10}, one can clearly observe the dark region inside the inner edge of the accretion disk, where the red regions correspond to the maximum temperature of the disk.

\section{Conclusions}
\label{Sec:Conclusion}
In this work, we studied the epicyclic motion of charged particles and the associated QPOs around the weakly magnetized BH surrounded by quintessence within the framework of Einstein-ModMax theory. The dynamics of charged particles moving on circular orbits have been analyzed, and the corresponding radial and vertical epicyclic frequencies are computed. In particular, we have been examining the influence of the nonlinear ModMax parameter, electromagnetic coupling, dyonic charge, and the quintessence state parameter on the ISCOand the characteristic frequencies governing particle motion in the strong-field regime.

Our result reveals that the electromagnetic charge and coupling parameters significantly affect the structure of circular orbits and the location of the ISCO. We have noted that an increase in the BH dyonic charge shifts the ISCO radius inward, accompanied by a decrease in the specific energy and angular momentum of the particles in the ISCO orbit, whereas the electric and magnetic coupling parameters modify the effective potential, and for the case $g_{e} > 0$, shift the ISCO radius outward. Although for the scenario $g_{e} < 0$, compressing the ISCO radius closer to the BH because the negative values of ge contribute to enhancing the gravitational force. Moreover, it is found that the quintessence parameter exerts only a weak influence on the ISCO radius, while the nonlinear ModMax parameter $\nu$ produces a more pronounced outward shift of the ISCO as its value is increased.

The behavior of the Keplerian frequency, orbital velocity, and the fundamental epicyclic frequencies was analyzed in detail. We demonstrated that the radial, vertical, and orbital frequencies exhibit distinct sensitivities to the model parameters in the strong-gravity region, while their asymptotic behavior at large radii remains consistent with the Keplerian scaling. The coupling parameter $g_{e}$ enhances radial oscillations while suppressing the vertical and azimuthal oscillations. In the case of magnetic coupling $\sigma_{m}$, we observe that the latitudinal frequency $\omega_{\theta}$ is affected by $\sigma_{m}$ only when $g_{e}=0.1$. In contrast, the orbital and radial frequencies remain unchanged.

Using the forced resonance model, we compared the theoretical predictions of high-frequency QPOs with observational data from several X-ray binary systems. We performed MCMC analysis to constrain the parameters of the weakly magnetized BH surrounded by quintessence in Einstein–ModMax theory. Our results reveal that QPO observations tightly constrain the BH mass and the orbital radius while placing stringent upper bounds on the ModMax $\nu$, coupling parameter $g_{e}$, the magnetic parameter $\sigma_{m}$, and the dimensionless charge $Q/M$. The observational data do not require significant deviations from standard Einstein-Maxwell theory, although a nonzero contribution from the quintessence parameter is mildly favored, indicating sensitivity to the surrounding quintessence field.

Finally, we have examined the radiative properties of the accretion disk in the vicinity of the considered BH. Our analysis showed that the energy flux $\mathcal{F}$, temperature, $T$, and energy $E=k\times T$ of the disk increase with the increasing dyonic charge, while they decrease with increasing values of the quintessence and ModMax parameters due to the weakening of the effective gravitational potential. Overall, our results indicate that accretion-disk properties and observed QPOs are consistent with GR in the strong-field regime, allowing only small deviations associated with weak magnetization and nonlinear electrodynamics in the presence of a quintessence background. 

\section*{Acknowledgments}

The research is supported by the National Natural Science Foundation of China under Grant No.~W2433018, No.~12275238, No.~12542053, and No. 11675143, the National Key Research and Development Program under Grant No.~2020YFC2201503, and the Zhejiang Provincial Natural Science Foundation of China under Grants No.~LR21A050001 and No.~LY20A050002, and the Fundamental Research Funds for the Provincial Universities of Zhejiang in China under Grant No. RF-A2019015.

%


\end{document}